\documentclass[lettersize,journal]{IEEEtran}
\usepackage{amsmath,amsfonts}
\usepackage{algorithmic}
\usepackage{algorithm}
\usepackage{array}
\usepackage[caption=false,font=normalsize,labelfont=sf,textfont=sf]{subfig}
\usepackage{textcomp}
\usepackage{stfloats}
\usepackage{url}
\usepackage{verbatim}
\usepackage{graphicx}

\usepackage{cite}
\usepackage{xcolor,colortbl}

\definecolor{Gray1}{gray}{0.85}
\definecolor{Gray2}{gray}{0.93}
\definecolor{LightCyan}{rgb}{0.88,1,1}
\newcolumntype{a}{>{\columncolor{Gray}}c}
\RequirePackage{booktabs}%

\begin{document}

\title{Analysis and Enhancement of Incremental-Quantity-Based Distance Protection With Grid-Forming Inverters}

%none or non

%\author{IEEE Publication Technology,~\IEEEmembership{Staff,~IEEE,}

%\author{Henrik Johansson,~\IEEEmembership{Staff,~IEEE}, Qianli Xing,~\IEEEmembership{Staff,~IEEE}, Nathaniel Taylor,~\IEEEmembership{Senior Member,~IEEE}, and Xiongfei Wang,~\IEEEmembership{Fellow,~IEEE}

\author{Henrik Johansson,~\IEEEmembership{Student Member,~IEEE}, Qianli Xing,~\IEEEmembership{Student Member,~IEEE}, Nathaniel Taylor,~\IEEEmembership{Senior Member,~IEEE}, and Xiongfei Wang,~\IEEEmembership{Fellow,~IEEE}

        % <-this % stops a space
%\thanks{This paper was produced by the IEEE Publication Technology Group. They are in Piscataway, NJ.}% <-this % stops a space
%\thanks{Manuscript received April 19, 2021; revised August 16, 2021.}}
\thanks{Henrik Johansson, Qianli Xing and Nathaniel Taylor are with the Department of Electric Power \& Energy Systems at KTH Royal Institute of Technology, SE-100 44 Stockholm, Sweden}
\thanks{Xiongfei Wang is with the Department of Electrical Engineering at Tsinghua University, 100084 Beijing, China}}

% The paper headers
%\markboth{Journal of \LaTeX\ Class Files,~Vol.~14, No.~8, August~2021}%
%{Shell \MakeLowercase{\textit{et al.}}: A Sample Article Using IEEEtran.cls for IEEE Journals}

%\IEEEpubid{0000--0000/00\$00.00~\copyright~2021 IEEE}
% Remember, if you use this you must call \IEEEpubidadjcol in the second
% column for its text to clear the IEEEpubid mark.

\maketitle

\begin{abstract}

Grid-forming (GFM) inverters are expected in future inverter-dominated grids. In such grids, time-domain protection schemes, for example those based on instantaneous incremental quantities (IQs), are being advocated as potential solutions to the challenges faced by traditional phasor-based protection schemes, due to their ability to process nonlinear data. However, IQ-based protection uses the superposition principle; thus, linearity is still assumed in their application, while GFM inverters are nonlinear sources during faults. This paper proposes an analytical model to study the impact of GFM inverters on the relay-measured IQs. The model is validated with PSCAD/EMTDC simulations, and is used to investigate the interoperability of time-domain IQ-based distance protection with GFM inverters employing different current limiters. Results show that time-domain IQ-based distance protection demonstrates superior dependability for close-in faults compared to that of quadrilateral distance protection with GFM inverters, and it has the possibility to be secure for external faults when quadrilateral distance protection overreaches; however, tuning of its settings is hard to generalize for various sources and faults. Taking the observed interoperability issues into account, a trip criterion for dependable and secure time-domain IQ-based distance protection is proposed, which facilitates easy-to-tune and general settings for applications with GFM inverters.

\end{abstract}


\section{Introduction}

\IEEEPARstart{W}{ith} the installed capacity of inverter-based resources (IBRs) increasing, such as wind and solar power, the fault characteristics of modern power grids keep evolving from being governed by the physical laws of rotating machines with high inertia and overcurrent capability, to being based on the implemented fault-ride through (FRT) control of the IBRs, which in turn depends on the adopted grid codes and how it is specifically tailored by the manufacturer to meet their own performance requirements \cite{Protection-100, IBR-impact-Haddadi, Distance-unconventional, IBR-impact-Juan, IBR-problems-Normann, GFM-ENTSO-E, IBR-impact-2025}. This development challenges conventional power system protection schemes, where the interoperability issues are generally related to the low inertia and fault current capability of IBRs, as well as their heterogeneous fault responses \cite{IBR-problems-Normann, Protection-100, Distance-unconventional}. The two former characteristics make IBRs nonlinear weak-infeed sources during faults, where the FRT control causes dynamical variations of their internal voltage source (IVS) and source impedance. Consequently, there are limitations to properly characterizing IBRs as fundamental frequency phasor Thévenin equivalent sources during faults---one of the most fundamental concepts underlying the vast majority of legacy protection schemes \cite{HJ-GFM-distance}. 

Traditional apparent impedance-based distance protection stands as one of the commonly employed protection schemes that is heavily impacted by the transition from synchronous generators (SGs) to IBRs in bulk power systems \cite{Protection-100, IBR-impact-Haddadi, Distance-unconventional, IBR-impact-Juan, Bergeron-protection-GFM, IBR-problems-Normann, GFM-ENTSO-E, IBR-impact-2025, Collaborative-distance-IBR}. Both dependability and security issues have been observed from field applications with lines connected to IBRs, such as zone 1 overreaching for external zone 2 faults and zone 1 missing or tripping late for close-in faults \cite{IQ-Naidu-souce-impedance, IQ-Naidu, Source-agnostic-double-line}. Similar issues have been observed through various simulation studies with grid-following (GFL) and grid-forming (GFM) inverters \cite{HJ-GFM-distance, GFM-distance-Baeckeland, GFM-distance-Booth, IBR-impact-Haddadi, IBR-problems-Normann, Collaborative-distance-IBR}. Dependability issues with zone 2 have also been observed, where it may drop out from operation due to it computing an oscillating apparent impedance \cite{HJ-GFM-distance, IBR-problems-Normann}. 

Moving to time-domain distance protection schemes is being advocated as a potential solution in prospective inverter-dominated grids \cite{IQ-Naidu, IQ-Vermunicht, Bergeron-protection-GFM, LCD-IQ-transient-energy}. For instance, differential equation algorithms (DEAs) can work with nonlinear voltage and current waveforms in the time domain to numerically estimate the line resistance and inductance to the fault point \cite{DEA-Type3, HJ-DEA-HongKong, Bergeron-protection-GFM}. In essence, DEAs are time-domain model-based algorithms, making the design of the filtering process crucial to obtain stable estimates of the line parameters and the distance to the fault point. However, such single-ended distance protection schemes still suffer from significant steady-state errors in case of weak-infeed conditions during resistive faults (model errors), which can be caused by the limited overcurrent capability of IBRs. 

%\IEEEpubidadjcol

Another distance protection alternative to the traditional one is the underreaching distance element based on time-domain incremental quantities (IQs) \cite{IQ-Naidu, IQ-real-world-faults, IQ-circle, IQ-Tanbhir, IQ-park}. Although time-domain IQs can work with nonlinear and nonstationary data, their application in distance protection relates them to a pure-fault network. The concept uses the superposition principle and assumes linear sources, although IBRs are nonlinear sources during faults. Therefore, time-domain IQ-based distance algorithms are affected by IBRs, to an extent that depends on the speed that dependable and secure relay decisions can be made, as well as on what way and how fast IBRs adjust their internal source parameters. Recently published papers have studied interoperability issues between nonlinear IBR FRT control dynamics and various IQ-based protection elements \cite{HJ-IQ-Bilbao, IQ-IBR-impact, Jianping-Bilbao, GFM-distance-Booth, IQ-Vermunicht}; however, they are case-specific and numerical studies, leaving the interoperability issues inconclusive. Hence, there is a need for a more generic modelling approach of the impacts that IBRs have on the relay-measured IQs, to better understand the performance advantages and disadvantages of IQ-based protection schemes over alternative ones. 

This paper proposes a general analytical framework to model the impact of GFM inverters on the relay-measured IQs. It allows for a qualitative and quantitative analysis of various IQ-based protection elements with GFM inverters (a similar modelling approach works with GFL inverters). The analytical model is used to thoroughly study the interoperability of time-domain IQ-based distance protection with GFM inverters employing different current-limiting control strategies. A performance comparison is conducted with quadrilateral distance protection near GFM inverters. The modelling approach is validated with PSCAD simulations on a simple test system and the analytical findings are validated with benchmark PSCAD simulations of two different GFM inverter control strategies. 

Results show that time-domain IQ-based distance protection has the possibility to achieve superior dependability for close-in faults compared to quadrilateral distance protection with GFM inverters, and has the possibility to be secure for external faults when quadrilateral distance protection overreaches; however, tuning of its settings is hard to generalize for various sources and fault cases. Therefore, a novel trip criterion for dependable and secure time-domain IQ-based distance protection is proposed, where time variables are used for tripping based on the shape of the running sum signals, instead of a trip threshold level. This facilitates easy-to-tune and straightforward settings for applications with both SGs and IBRs, including GFM inverters with different current-limiting control strategies. The proposed trip criterion is validated with PSCAD simulations of the two implemented GFM inverters.

Following this introduction, Section~\ref{sec:system-description} briefly describes the studied system and the GFM inverter models implemented in PSCAD to conduct the simulation studies. Section~\ref{sec:Fundamentals-IQ} explains the fundamentals of time-domain IQ-based distance protection, after which  Section~\ref{sec:Proposed-modelling} describes the proposed analytical framework to model the impact of GFM inverters on the relay-measured IQs and IQ-based distance protection. Section~\ref{sec:Interoperability-IQ-dist-GFM} evaluates the interoperability of IQ-based distance protection and quadrilateral distance protection with GFM inverters, and the novel trip criterion for time-domain IQ-based distance protection is proposed and validated. Lastly, Section~\ref{sec:Conclusion} concludes the findings of the paper.

\section{System Description}
\label{sec:system-description}
\begin{figure}[h]
    \centering
    \includegraphics[width=1\linewidth]{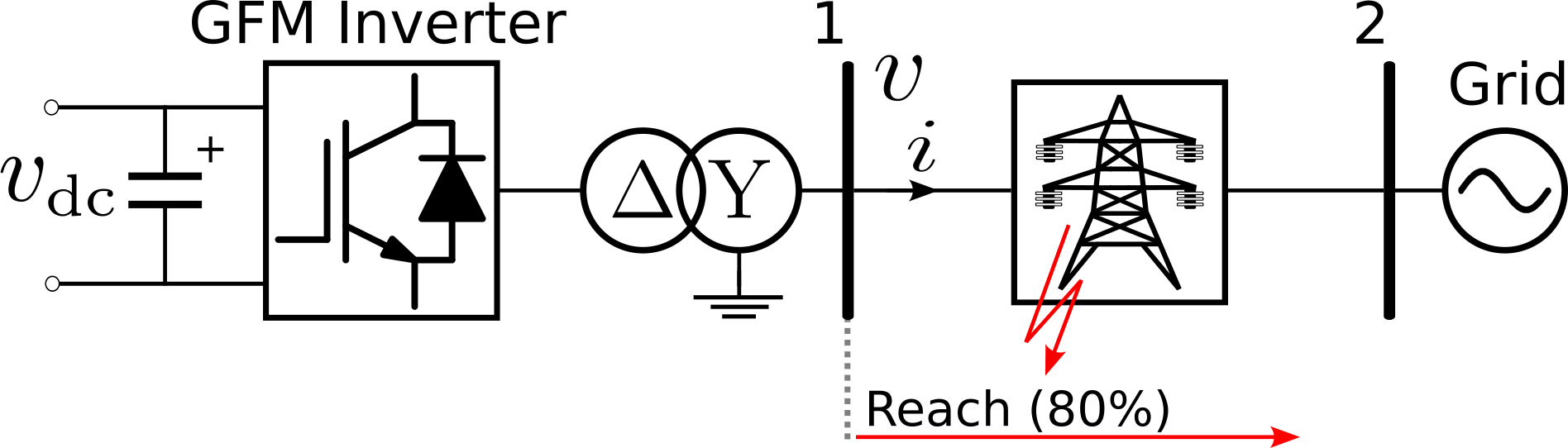}
    \caption{Studied power system.}
    \label{fig:SLIB-system}
\end{figure}

Fig.~\ref{fig:SLIB-system} shows the studied power system. It includes a $66$~kV GFM inverter at the sending end (left side) and a $\Delta$-Y$0$ transformer with negligible leakage reactance that raises the voltage to $220$ kV. The system frequency is $50$ Hz and the nominal power of the GFM inverter is $300$ MW. A $100$~km transposed overhead line (OHL) transfers the power to the utility grid and is modelled using the frequency-dependent (phase) model available in PSCAD. The remote-end source has a source-to-line impedance ratio (SIR) of $0.3$ (typical for long lines \cite{SIR-tutorial}) and the GFM inverter output filter also yields a SIR of $0.3$. The analyzed distance relay is located at the terminal of the sending-end source and has a reach of $80\%$ of the total line length. 

\begin{figure}[h]
    \centering
    \includegraphics[width=1\linewidth]{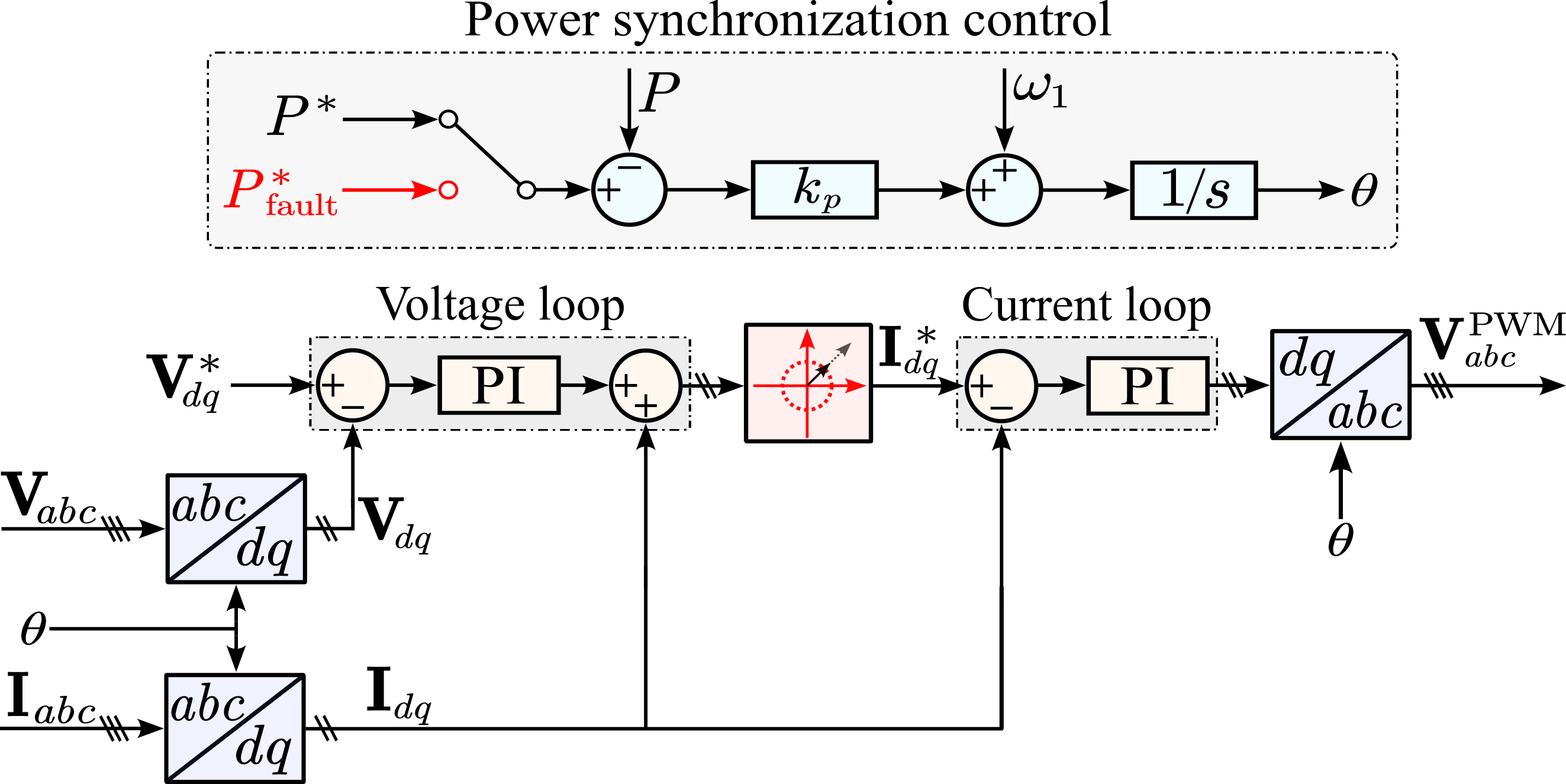}
    \caption{GFM1 control diagram as shown in \cite{HJ-GFM-distance}.}
    \label{fig:GFM1-schematic}
\end{figure}

\begin{figure}[h]
    \centering
    \includegraphics[width=1\linewidth]{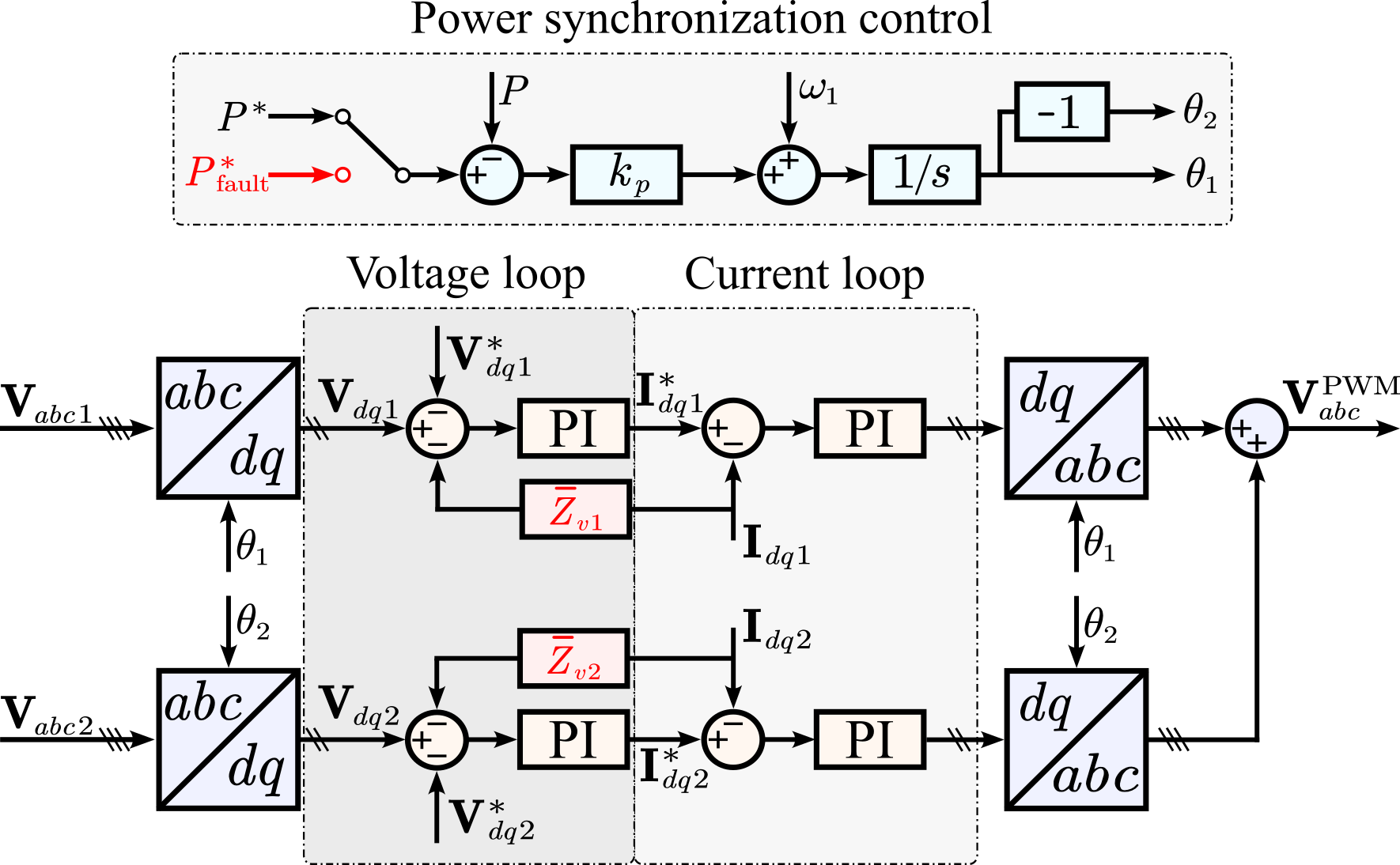}
    \caption{GFM2 control diagram as shown in \cite{HJ-GFM-distance}.}
    \label{fig:GFM2-schematic}
\end{figure}

Two different GFM inverter control schemes are implemented in the system shown in Fig.~\ref{fig:SLIB-system} (one at a time) to validate the analytical findings and the proposed IQ-based distance element. They have been explained in \cite{HJ-GFM-distance} and will in this paper be referred to as GFM1 and GFM2, respectively. Fig.~\ref{fig:GFM1-schematic} and \ref{fig:GFM2-schematic} show their control diagrams. Both employ the $P$-$f$ droop power synchronization control (PSC) and have an inductive output filter, but unlike GFM1, GFM2 separately controls the positive- and negative-sequences. Moreover, their current-limiting control strategies are vastly different: GFM1 applies a $1.2$~pu magnitude saturation current limiter (no angle prioritization), while GFM2 applies an adaptive virtual impedance to curtail the injected current by reducing the voltage reference going into the outer voltage control loop \cite{HJ-GFM-distance}. The angle of the GFM2 virtual impedance is the same as that of the positive-sequence line impedance (mainly inductive). 

Thorough comparisons of the current limiters implemented in GFM1 and GFM2 are given in \cite{GFM-current-limitation-Xiongfei, GFM-current-limitation-Baeckeland}. In short, the high control bandwidth of the inner current control loop enables the saturation current limiter of GFM1 to quickly limit the injected current to the desired value, providing superior current-limiting dynamics over the current-limiting control employed in GFM2. However, from the power system point of view, the virtual-impedance-based current limiter of GFM2 emulates an additional source impedance with the magnitude and angle used in the current-limiting controller (can at first be adaptive and then get a fixed value), whereas the saturation current limiter of GFM1 that keeps the angle of the current reference output from the outer voltage control loop, emulates an adaptive virtual resistance \cite{GFM-limiter-sat-resistive}.

This paper focuses on balanced faults and all the simulated fault cases with either GFM1, GFM2 or an SG are:
\begin{itemize}
    \item Balanced ABCG faults
    \item Distance from the relay (normalized) \newline $m_\mathrm{f}\in \{0.2, 0.45, 0.5, 0.6, 0.7, 0.75, 0.81, 0.9, 0.99 \}$ 
    \item Fault resistance $R_\mathrm{f} \in \{0, 2, 5, 8, 10, 15\}$ $\Omega$
    \item Fault inception angle $\phi_\mathrm{f} =0^\circ$ (always balanced faults)
\end{itemize}
and the pre-fault load condition is set to $1$~pu active power. This results in $54$ balanced fault simulations with each source. Interoperability between IQ-based distance protection and GFM inverters have been found to be particularly challenging for balanced faults \cite{HJ-IQ-Bilbao}.

\section{Fundamentals of Time-Domain IQ-Based Distance Protection}
\label{sec:Fundamentals-IQ}

The relay at bus $1$ in Fig.~\ref{fig:SLIB-system} measures the sending-end voltages $v_{\mathrm{s}x}(t)$ and currents $i_{\mathrm{s}x}(t)$, where $x\in \{\mathrm{a, b, c}\}$ is any phase. A memory function can be used to compute the line-to-ground (LG) voltage and current IQs according to
\begin{equation}
    \begin{aligned}
        &\Delta v_{\mathrm{s}x}(t) = v_{\mathrm{s}x}(t) - v_{\mathrm{s}x}(t-pT) \\
        &\Delta i_{\mathrm{s}x}(t) = i_{\mathrm{s}x}(t) - i_{\mathrm{s}x}(t-pT),
    \end{aligned}
    \label{IQ-definition}
\end{equation}
where $T= 20$ ms is the fundamental period and $p$ is an integer number. The sending-end line-to-line (LL) voltages and currents as well as their corresponding IQs are also computed. The purpose of IQ-based distance protection is not to directly estimate the distance to the actual fault point $m_\mathrm{f}$, but to ensure the relay only trips for faults up to the relative line length $m$ (reach). For each of the six fault loops, its fundamental principle is based on computing the absolute value of the voltage change between a certain number of power cycles at the reach point (operating quantity), and compare it with the absolute value of the pre-fault voltage at the reach point (restraining quantity) \cite{IQ-circle, IQ-Tanbhir, IQ-real-world-faults}. 

\begin{figure}[h]
    \centering
    \includegraphics[width=0.8\linewidth]{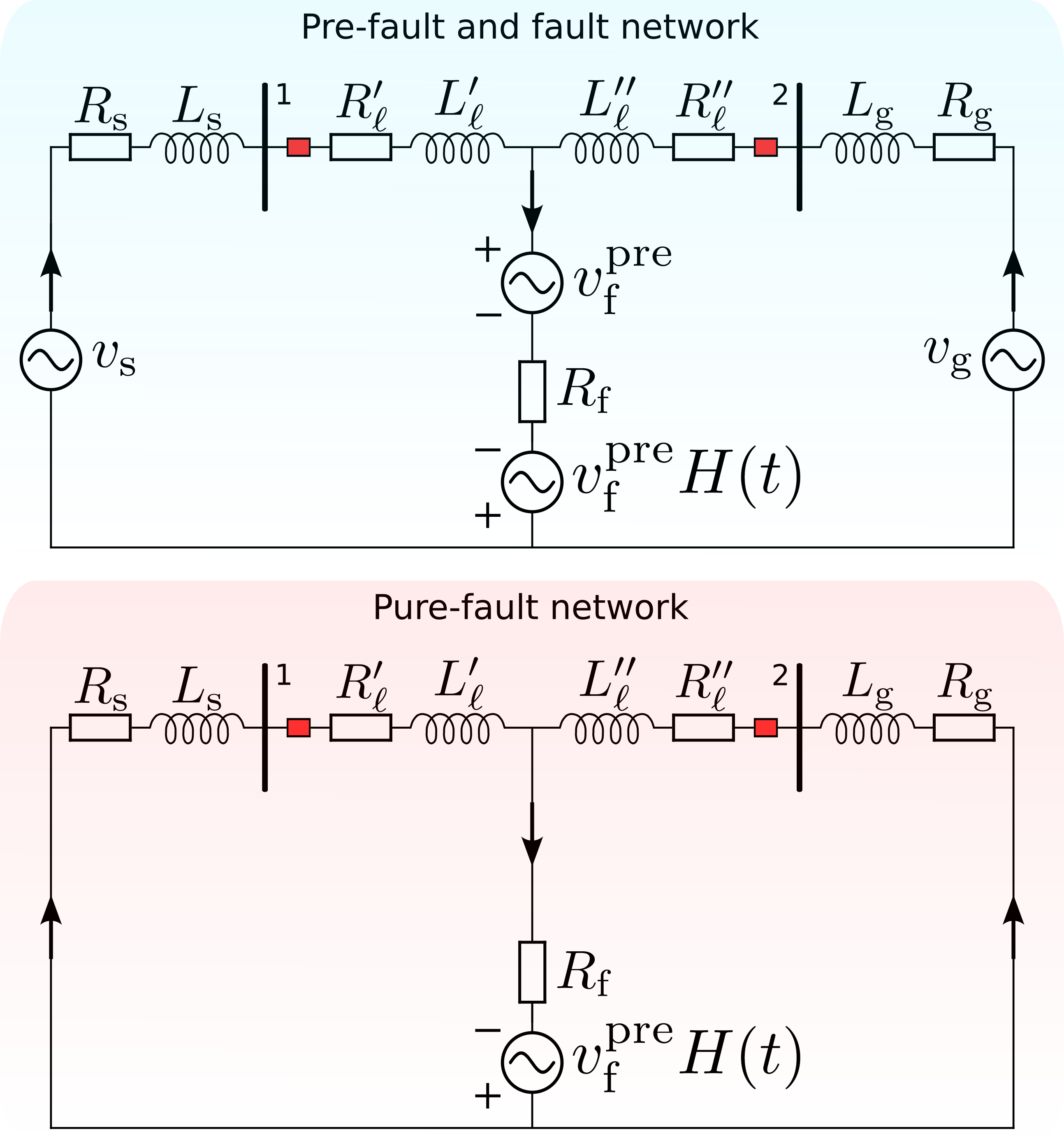}
    \caption{Time-domain pure-fault network under balanced faults (linear sources).}
    \label{fig:Pure-fault-TD}
\end{figure}

In SG-dominated power systems, the operating quantity of a faulted loop will generally be larger in magnitude than the restraining quantity for the majority of the data samples when the fault is within the reach of the relay ($m_\mathrm{f}<m$). An external fault would cause the opposite to occur \cite{IQ-circle}. However, a high fault resistance $R_\mathrm{f}$ could jeopardize these properties. Moreover, the IQ-based distance protection principle assumes the sending-end source is linear to construct the so-called ``pure-fault'' network---an assumption prone to errors since it neglects the nonlinear sending-end IBR source dynamics in Fig.~\ref{fig:SLIB-system}. The pure-fault network is constructed using the superposition principle, as shown in Fig.~\ref{fig:Pure-fault-TD}, where $H(t)$ is the Heaviside step function that triggers the fault. With SGs (linear sources), the IQs (\ref{IQ-definition}) reside within the pure-fault network. 

\subsection{Operating and Restraining Quantities}
\label{subsubsec: dist-IQ-op-rst}
Solid faults are assumed, the stray capacitance in the line model is neglected and the line is considered to be ideally transposed when acquiring the operating quantities of all six fault loops. Uppercase letters will be used to denote the values of the line parameters over its entire length. In that case, the incremental voltage drop over the line to the reach point $m$ is computed according to
\begin{equation}
\begin{aligned}
    \Delta v_{mx} &= mR_1\Delta i_{\mathrm{s}x} + m(R_0-R_1)\Delta i_{\mathrm{s}0} \\
    &+ mL_1\frac{d}{dt}\Delta i_{\mathrm{s}x} + m(L_0-L_1)\frac{d}{dt}\Delta i_{\mathrm{s}0}\\
    \Delta v_{mxy} &= mR_1\Delta i_{\mathrm{s}xy} + mL_1\frac{d}{dt}\Delta i_{\mathrm{s}xy},
\end{aligned} 
\label{IQ-v-drop}
\end{equation}
where $x \in \{\mathrm{a}, \mathrm{b},\mathrm{c} \}$ is any phase, $y$ is the subsequent phase in the positive phase sequence and $\Delta i_\mathrm{s0} = (\Delta i_\mathrm{sa}+\Delta i_\mathrm{sb}+\Delta i_\mathrm{sc})/3$ is the instantaneous zero-sequence incremental current. Using (\ref{IQ-v-drop}), the LG operating quantities $\psi_{x}^\mathrm{op}$ and LL operating quantities $\psi_{xy}^\mathrm{op}$, can be obtained:
\begin{equation}
    \begin{aligned}
        \psi_{x}^\mathrm{op} &= |\Delta v_{\mathrm{s}x} - \Delta v_{mx}| \\
        \psi_{xy}^\mathrm{op} &= |\Delta v_{\mathrm{s}xy} - \Delta v_{mxy}|, 
    \end{aligned}
    \label{Operating}
\end{equation}
where $\Delta v_{\mathrm{s}x}$ and $\Delta v_{\mathrm{s}xy}$ are the LG and LL sending-end incremental voltages measured by the relay, respectively. 

All six restraining quantities can be acquired in a similar way as the operating quantities but not using the IQs measured by the sending-end relay. However, memory must be used to store the pre-fault voltage at the reach point. First, the LG and LL fault loop voltage drops over the transmission line to the reach point are computed:
\begin{equation}
\begin{aligned}
    v_{mx} &= mR_1 i_{\mathrm{s}x} + m(R_0-R_1)i_{\mathrm{s}0} \\
    &+ mL_1\frac{d}{dt} i_{\mathrm{s}x} + m(L_0-L_1)\frac{d}{dt} i_{\mathrm{s}0}\\
    v_{mxy} &= mR_1 i_{\mathrm{s}xy} + mL_1\frac{d}{dt} i_{\mathrm{s}xy}. 
\end{aligned} 
\label{v-drop}
\end{equation}
Then, the LG restraining quantities $\psi_{x}^\mathrm{rst}$ and LL restraining quantities $\psi_{xy}^\mathrm{rst}$, are computed according to
\begin{equation}
    \begin{aligned}
        \psi_{x}^\mathrm{rst} &= K|v_{\mathrm{s}x}(t-pT) - v_{mx}(t-pT)| \\
        \psi_{xy}^\mathrm{rst} &= K|v_{\mathrm{s}xy}(t-pT) - v_{mxy}(t-pT)|, 
    \end{aligned}
    \label{Restraining}
\end{equation}
where $K\geq1$ is a constant, $T=20$~ms is the fundamental period and $p$ is the same integer number used to compute the voltage and current IQs in (\ref{IQ-definition}). The factor $K\geq1$ can be applied to slightly bias the relay towards security, at the expense of less dependability \cite{IQ-park, IQ-real-world-faults}. Notice that multiplying the restraining quantities with a factor $K\geq1$ is different to simply decreasing the reach $m$. The derivative terms in (\ref{IQ-v-drop}) and (\ref{v-drop}) can be numerically computed using the method of central finite difference.

\subsection{Trip Variables}
\label{sec:IQ-dist-trip-variables}

With strong conventional sources, solid faults occurring before the designated reach point $m$ (internal) cause the operating quantities of the faulted loops to be larger in magnitude than their respective restraining quantities for most data samples. This condition is not instantaneous because the operating quantities develop from zero while the restraining quantities do not; yet, it is commonly adopted for tripping (with some variations) in time-domain IQ-based distance protection \cite{IQ-circle, IQ-Tanbhir, IQ-real-world-faults, IQ-park}. The time-domain implementation is often based on computing the integral of the difference between the operating quantities and the corresponding restraining quantities:
\begin{equation}
    \begin{aligned}
        &E_{x}(t) = \int_{t_1}^{t} \Big( \psi_{x}^\mathrm{op}(\tau) - \psi_{x}^\mathrm{rst}(\tau) \Big) \hspace{0.05cm} d\tau \\
        &E_{xy}(t) = \int_{t_1}^{t} \Big( \psi_{xy}^\mathrm{op}(\tau) - \psi_{xy}^\mathrm{rst}(\tau) \Big) \hspace{0.05cm} d\tau,
    \end{aligned}
\end{equation}
where $\tau$ is the time variable of integration, $\tau = t_1$ is the moment in time when the integration starts and $\tau = t$ is the latest moment in time (present). In discrete form, the integrals are replaced by running sums based on the short-window trapezoidal rule, partitioned into equally spaced subintervals of the sampling time period $T_\mathrm{s}$, according to
\begin{equation}
    \begin{aligned}
        E_{x}^\Sigma[n] = 0.5T_\mathrm{s}\sum_{k=k_1}^{n} &\Big( \psi_{x}^\mathrm{op}[k] + \psi_{x}^\mathrm{op}[k-1] \\ &- \psi_{x}^\mathrm{rst}[k] - \psi_{x}^\mathrm{rst}[k-1] \Big) \\
        E_{xy}^\Sigma[n] = 0.5T_\mathrm{s}\sum_{k=k_1}^{n} &\Big( \psi_{xy}^\mathrm{op}[k] + \psi_{xy}^\mathrm{op}[k-1] \\ &- \psi_{xy}^\mathrm{rst}[k] - \psi_{xy}^\mathrm{rst}[k-1] \Big), 
    \end{aligned}
    \label{Running-sum-equation}
\end{equation}
where $k_1$ is the first data sample when starting a specific running sum and $n$ is the latest (current) sample index available for the operating and restraining quantities.

The cumulative sum (\ref{Running-sum-equation}) tends to be an increasing value of a faulted loop during internal faults, while the opposite occurs under external faults and for the healthy loops under internal fault conditions. These features can be exploited to generate trip variables that discriminate between internal and external forward faults, where the trends of $E_{x}^\Sigma$ and $E_{xy}^\Sigma$ carry the main information for tripping. If a running sum in (\ref{Running-sum-equation}) exceeds a certain threshold, a trip signal may be issued \cite{IQ-circle, IQ-Tanbhir, IQ-real-world-faults, IQ-park}. Although additional security checks from other IQ-based protection elements may be implemented (fault detection, phase selection and directionality checking), this is the primary trip criterion used in time-domain IQ-based distance protection.

\section{Modelling of Nonlinear GFM Inverter FRT Dynamics on IQ-Based Protection Schemes}
\label{sec:Proposed-modelling}
To better understand the impacts that IBRs have on IQ-based protection schemes, the relay operation must be linked to the fault response of IBRs. In this section, such a generic model is proposed for GFM inverters, which extends the application of the superposition principle in Fig.~\ref{fig:Pure-fault-TD}. The pre-disturbance and disturbance network as well as the IQ network, are introduced as replacements for the pre-fault and fault network as well as the pure-fault network, respectively.

\subsection{Impact of GFM Inverter Nonlinear FRT Dynamics on the Relay-Measured IQs}
\label{GFM-IQ-impact-explain}
\begin{figure*}[ht]
    \centering
    \includegraphics[width=1\linewidth]{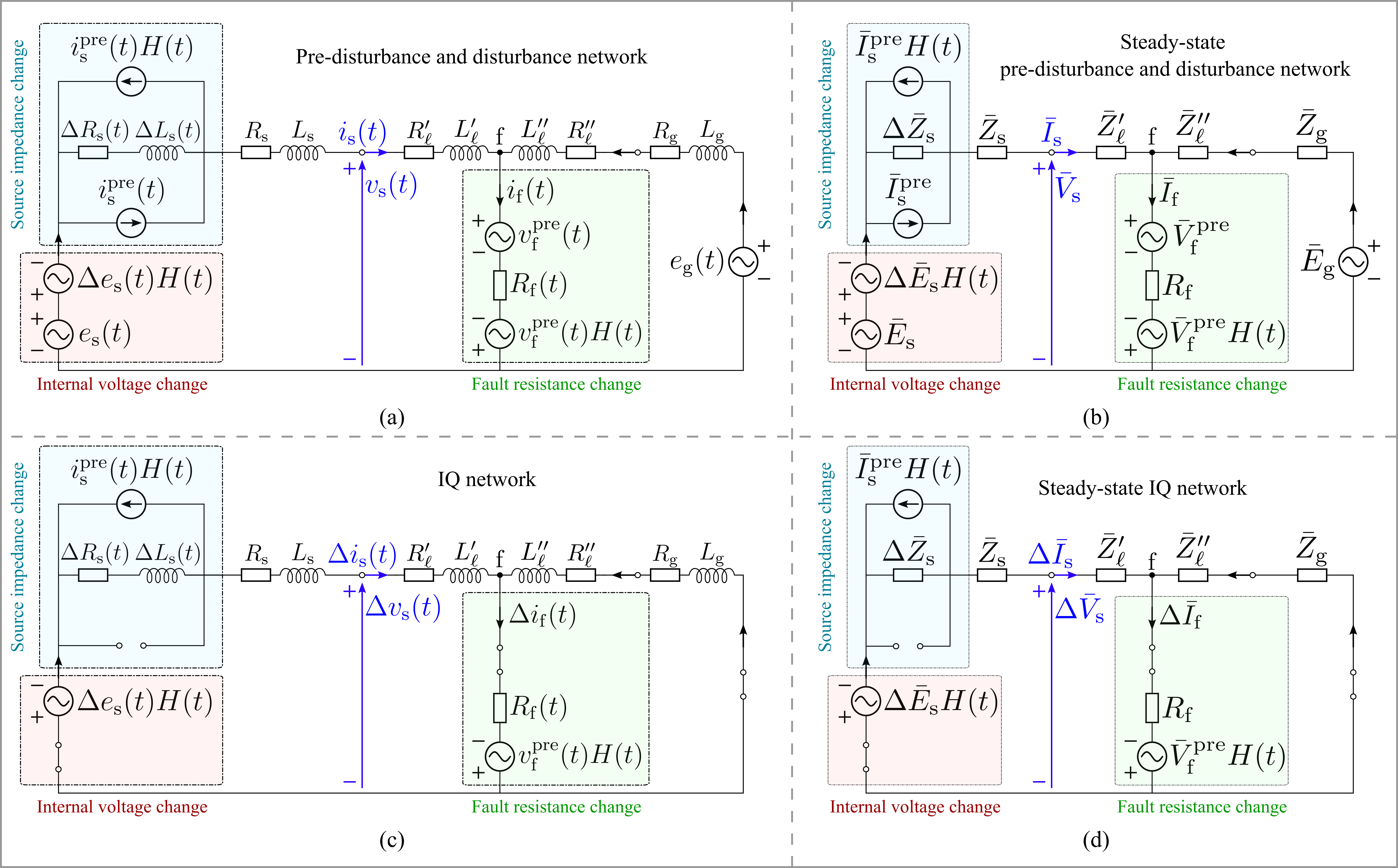}
    \caption{Proposed modelling of nonlinear GFM inverter FRT dynamics on the IQs locally measured by the relay during balanced faults.}
    \label{fig:IQ-network-TD-and-phasor}
\end{figure*}

From the power system point of view, limiting the injected current can be accomplished in two ways by a GFM inverter: (i) changing the IVS and/or (ii) increasing the source impedance. In Fig.~\ref{fig:IQ-network-TD-and-phasor}~(a), these dynamical source changes are modelled with the introduction of a variable voltage source $\Delta e_\mathrm{s}(t)H(t)$ as well as a variable resistance $\Delta R_\mathrm{s}(t)$ and inductance $\Delta L_\mathrm{s}(t)$. The function of time indicates that the GFM inverter current limitation is not instantaneous, but rather a dynamical process (also reactive power injection during resistive faults and synchronization). For generality, the fault resistance $R_\mathrm{f}(t)$ is also modelled as function of time. 

Note that the GFM inverter source impedance change is modelled using two current sources in parallel with the additional impedance, where $i_\mathrm{s}^\mathrm{pre}(t)$ is the pre-fault load current being injected by the GFM inverter. Kirchhoff's current law in both nodes connecting to $\Delta R_\mathrm{s}(t)$ and $\Delta L_\mathrm{s}(t)$, yields that no current is flowing through them prior to the fault inception. Once a fault occurs, the two $i_\mathrm{s}^\mathrm{pre}(t)$ current sources cancel in a similar way as the two $v_\mathrm{f}^\mathrm{pre}(t)$ voltage sources cancel at the fault inception. The extreme case when either $\Delta R_\mathrm{s}(t)\rightarrow\infty$ or $\Delta L_\mathrm{s}(t)\rightarrow\infty$, results in $i_\mathbf{s}(t)\rightarrow 0$ and $\Delta i_\mathrm{s}(t)\rightarrow - i_\mathrm{s}^\mathrm{pre}(t)$. 

For the purpose of limiting the injected current, changes of both the IVS and source impedance may be considered simultaneously; however, it is sufficient to only consider one of them. For example, in the case of an adaptive virtual impedance for current limitation, $\Delta R_\mathrm{s}(t)$ and $\Delta L_\mathrm{s}(t)$ could be used to model the current limiter directly, whereas $\Delta e_\mathrm{s}(t)$ could instead be used to model it indirectly by representing the time-varying voltage drop over the adaptive virtual impedance. If instead the IVS magnitude is dynamically reduced to limit the injected current, $\Delta e_\mathrm{s}(t)$ models it directly from the power system point of view, whereas $\Delta R_\mathrm{s}(t)$ and $\Delta L_\mathrm{s}(t)$ would instead model it indirectly, by representing the time-varying resistance and inductance whose collective voltage drop yields the variable difference between the pre-fault IVS and the IVS during the fault.

\subsection{Produced IQs due to GFM Inverters}
The phasor-based steady-state equivalent of Fig.~\ref{fig:IQ-network-TD-and-phasor}~(a) is shown in Fig.~\ref{fig:IQ-network-TD-and-phasor}~(b). By applying the superposition principle to remove the pre-disturbance load data in Fig.~\ref{fig:IQ-network-TD-and-phasor}~(a), the time-domain IQ network is obtained, as shown in Fig.~\ref{fig:IQ-network-TD-and-phasor}~(c), whose steady-state equivalent becomes as in Fig.~\ref{fig:IQ-network-TD-and-phasor}~(d). For brevity and ease of analysis, the steady-state IQ network will be used to derive the expressions of the sending-end IQs measured by the relay and its performance. This will be motivated further in Section~\ref{Sec:Sim-validation}. Moreover, the calculations will be carried out per phase, since only balanced disturbances are considered. 

In order to solve for $\Delta\bar{I}_\mathrm{s}$ and $\Delta\bar{V}_\mathrm{s}$ in Fig.~\ref{fig:IQ-network-TD-and-phasor}~(d), it is practical to once again apply the superposition principle to construct three separate circuit diagrams for each source energizing the IQ network: one for $\bar{I}_\mathrm{s}^\mathrm{pre}H(t)$, $\Delta\bar{E}_\mathrm{s}H(t)$ and $\bar{V}_\mathrm{f}^\mathrm{pre}H(t)$, respectively. When doing so, it becomes handy to define the following variables:
\begin{equation}
\begin{aligned}
    &\bar{\mathbb{Z}}_\mathrm{x}= (\bar{Z}_\mathrm{s}+\Delta\bar{Z}_\mathrm{s} + m_\mathrm{f}\bar{Z}_\mathrm{\ell})(\bar{Z}_\mathrm{g} + (1-m_\mathrm{f})\bar{Z}_\mathrm{\ell}) \\
    & \bar{Z}_\mathrm{sg\ell} = \bar{Z}_\mathrm{s} + \Delta\bar{Z}_\mathrm{s} + \bar{Z}_\mathrm{g} + \bar{Z}_\mathrm{\ell} \\
    &\bar{Z}_\mathrm{y} = R_\mathrm{f}||(\bar{Z}_\mathrm{g}+(1-m_\mathrm{f})\bar{Z}_\mathrm{\ell}),
\end{aligned}
\end{equation}
where $\bar{\mathbb{Z}}_\mathrm{x}$ has the unit $\Omega^2$ and has therefore been given a slightly different notation compared to an impedance. The sending-end incremental current measured by the relay is found from Fig.~\ref{fig:IQ-network-TD-and-phasor}~(d) to equal
\begin{equation}
\begin{aligned}
    \Delta\bar{I}_\mathrm{s} &= \bar{V}_\mathrm{f}^\mathrm{pre}\frac{\bar{\mathbb{Z}}_\mathrm{x}}{(\bar{Z}_\mathrm{s}+\Delta\bar{Z}_\mathrm{s} + m_\mathrm{f}\bar{Z}_\mathrm{\ell})(\bar{\mathbb{Z}}_\mathrm{x}+R_\mathrm{f}\bar{Z}_\mathrm{sg\ell})} \\
    &- \frac{1}{\bar{Z}_\mathrm{y}+m_\mathrm{f}\bar{Z}_\mathrm{\ell} + \bar{Z}_\mathrm{s}+\Delta\bar{Z}_\mathrm{s}}(\Delta\bar{E}_\mathrm{s}+\bar{I}_\mathrm{s}^\mathrm{pre}\Delta\bar{Z}_\mathrm{s}),
\end{aligned} 
\label{IQ-I-phasor}
\end{equation}
and the sending-end IQ voltage at the relay becomes
\begin{equation}
\begin{aligned}
    \Delta\bar{V}_\mathrm{s} &= -\bar{V}_\mathrm{f}^\mathrm{pre}\frac{(\bar{Z}_\mathrm{s}+\Delta\bar{Z}_\mathrm{s})\bar{\mathbb{Z}}_\mathrm{x}}{(\bar{Z}_\mathrm{s}+\Delta\bar{Z}_\mathrm{s} + m_\mathrm{f}\bar{Z}_\mathrm{\ell})(\bar{\mathbb{Z}}_\mathrm{x}+R_\mathrm{f}\bar{Z}_\mathrm{sg\ell})} \\
    &- \frac{\bar{Z}_\mathrm{y}+m_\mathrm{f}\bar{Z}_\mathrm{\ell}}{\bar{Z}_\mathrm{y}+m_\mathrm{f}\bar{Z}_\mathrm{\ell} + \bar{Z}_\mathrm{s}+\Delta\bar{Z}_\mathrm{s}}(\Delta\bar{E}_\mathrm{s}+\bar{I}_\mathrm{s}^\mathrm{pre}\Delta\bar{Z}_\mathrm{s}).
\end{aligned} 
\label{IQ-V-phasor}
\end{equation}
Note that even though the Heaviside step functions are activated at $t=0$, no IQs are produced if $R_\mathrm{f} \rightarrow \infty$, $\Delta\bar{E}_\mathrm{s}\rightarrow0$ and $\Delta\bar{Z}_\mathrm{s}\rightarrow 0$. This also applies to the time-domain IQ network in Fig.~\ref{fig:IQ-network-TD-and-phasor}~(c). In fact, $R_\mathrm{f}(t)$, $\Delta R_\mathrm{s}(t)$, $\Delta L_\mathrm{s}(t)$ and $\Delta e_\mathrm{s}(t)$ may take on any functions of time (discrete or continuous) and be activated at different points in time---the proposed IQ network in Fig.~\ref{fig:IQ-network-TD-and-phasor}~(c) still applies if the memory function in (\ref{IQ-definition}) remains within the pre-disturbance network, i.e. the IQs are still referred to the system changes from the initial load condition. Therefore, the relay-measured incremental voltages and currents reside within the circuit diagram shown in Fig.~\ref{fig:IQ-network-TD-and-phasor}~(c), whose steady-state equivalent becomes as in Fig.~\ref{fig:IQ-network-TD-and-phasor}~(d), when a GFM inverter is the sending-end source.

\subsection{Operating Quantities of IQ-Based Distance Protection due to GFM Inverters}
From the expressions in (\ref{IQ-I-phasor}) and (\ref{IQ-V-phasor}), the robustness of all IQ-based protection elements against nonlinear GFM inverter fault response characteristics can be evaluated in steady-state conditions. For the operating quantity used in IQ-based distance protection (\ref{Operating}), the following expression is obtained:
\begin{equation}
    \begin{aligned}
        \psi^\mathrm{op} &= \Bigg| -\bar{V}_\mathrm{f}^\mathrm{pre}\frac{(\bar{Z}_\mathrm{s}+\Delta\bar{Z}_\mathrm{s})\bar{\mathbb{Z}}_x + m\bar{Z}_\mathrm{\ell}\bar{\mathbb{Z}}_x}{(\bar{Z}_\mathrm{s}+\Delta\bar{Z}_\mathrm{s} + m_\mathrm{f}\bar{Z}_\mathrm{\ell})(\bar{\mathbb{Z}}_\mathrm{x} + R_\mathrm{f}\bar{Z}_\mathrm{sg\ell})} \\
        &- (\Delta\bar{E}_\mathrm{s} + \bar{I}_\mathrm{s}^\mathrm{pre}\Delta\bar{Z}_\mathrm{s})\frac{\bar{Z}_\mathrm{y}+m_\mathrm{f}\bar{Z}_\mathrm{\ell}-m\bar{Z}_\mathrm{\ell}}{\bar{Z}_\mathrm{y}+m_\mathrm{f}\bar{Z}_\mathrm{\ell}+\bar{Z}_\mathrm{s}+\Delta\bar{Z}_\mathrm{s}}\Bigg| \\
        &= \big|\bar{\psi}^\mathrm{op}\big|,
    \end{aligned}
    \label{Distance-IQ-op-GFM}
\end{equation}
which in the ideal case with a linear sending-end voltage source and a solid fault, becomes equal to
\begin{equation}
    \psi^\mathrm{op,ideal} = \Bigg|-\bar{V}_\mathrm{f}^\mathrm{pre}\frac{\bar{Z}_\mathrm{s}+m\bar{Z}_\mathrm{\ell}}{\bar{Z}_\mathrm{s}+m_\mathrm{f}\bar{Z}_\mathrm{\ell}}\Bigg| = \big|\bar{\psi}^\mathrm{op,ideal}\big|.
    \label{Distance-op-IQ-ideal}
\end{equation}
The restraining quantity is 
\begin{equation}
    \psi^\mathrm{rst} = \Big|\bar{V}_\mathrm{f}^\mathrm{pre}\Big| = \big|\bar{\psi}^\mathrm{rst} \big|,
    \label{Distance-IQ-rst}
\end{equation}
and it holds that 
\begin{itemize}
    \item $\psi^\mathrm{op,ideal} = \psi^\mathrm{rst} \Longrightarrow m_\mathrm{f} = m$
    \item $\psi^\mathrm{op,ideal} > \psi^\mathrm{rst} \Longrightarrow m_\mathrm{f} < m$ (internal fault)
    \item $\psi^\mathrm{op,ideal} < \psi^\mathrm{rst} \Longrightarrow m_\mathrm{f} > m$ (external fault).
\end{itemize}
This is a steady-state proof of the IQ-based distance protection operating principle explained in Section~\ref{sec:Fundamentals-IQ}, under balanced faults (it also holds for unbalanced faults). 

The complex functions $\bar{\psi}^\mathrm{op}$ (\ref{Distance-IQ-op-GFM}) and $\bar{\psi}^\mathrm{rst}$ (\ref{Distance-IQ-rst}) are in phase if the power system is linear and the initial system impedances have the same phase angles (homogeneous). However, when a resistive fault occurs, the magnitude of $\bar{\psi}^\mathrm{op}$ is reduced and its phase angle becomes different compared to that of $\bar{\psi}^\mathrm{rst}$. Moreover, when the sending-end source is a GFM inverter, the operating quantity in (\ref{Distance-IQ-op-GFM}) may be substantially different in both magnitude and phase. A lower magnitude of $\bar{\psi}^\mathrm{op}$ will negatively impact the dependability of IQ-based distance elements, both time-domain and steady-state implementations, whereas a phase shift between $\bar{\psi}^\mathrm{op}$ and $\bar{\psi}^\mathrm{rst}$, adds a superimposed second-order harmonic on the trend of the running sums (\ref{Running-sum-equation}) used for time-domain IQ-based distance protection, which further impacts both dependability and security, depending on where the trip threshold level is implemented. 

\subsection{Simulation Validation}
\label{Sec:Sim-validation}
\begin{figure}[h]
    \centering
    \includegraphics[width=1\linewidth]{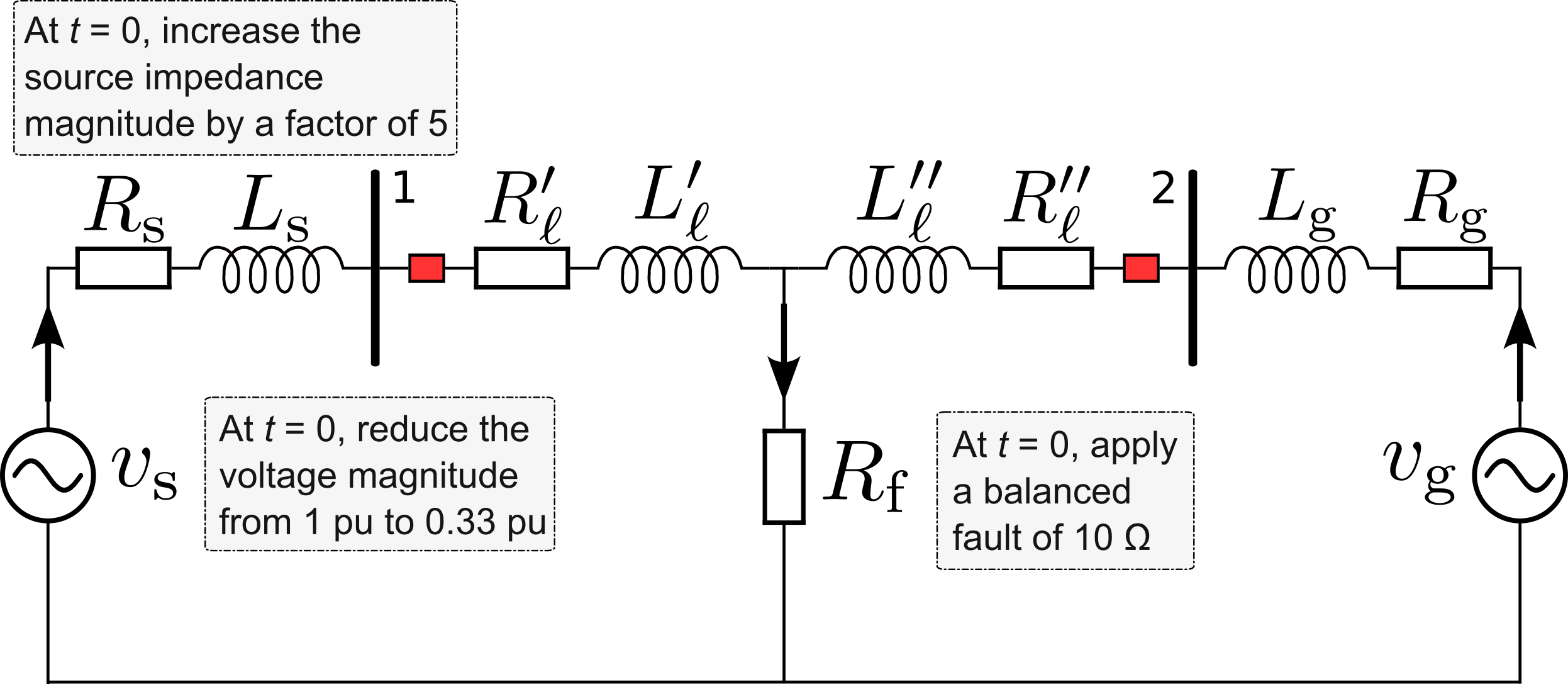}
    \caption{PSCAD test system for simulation validation.}
    \label{fig:IQ-PSCAD-setup}
\end{figure}

\begin{figure}[h]
    \centering
    \includegraphics[width=1\linewidth]{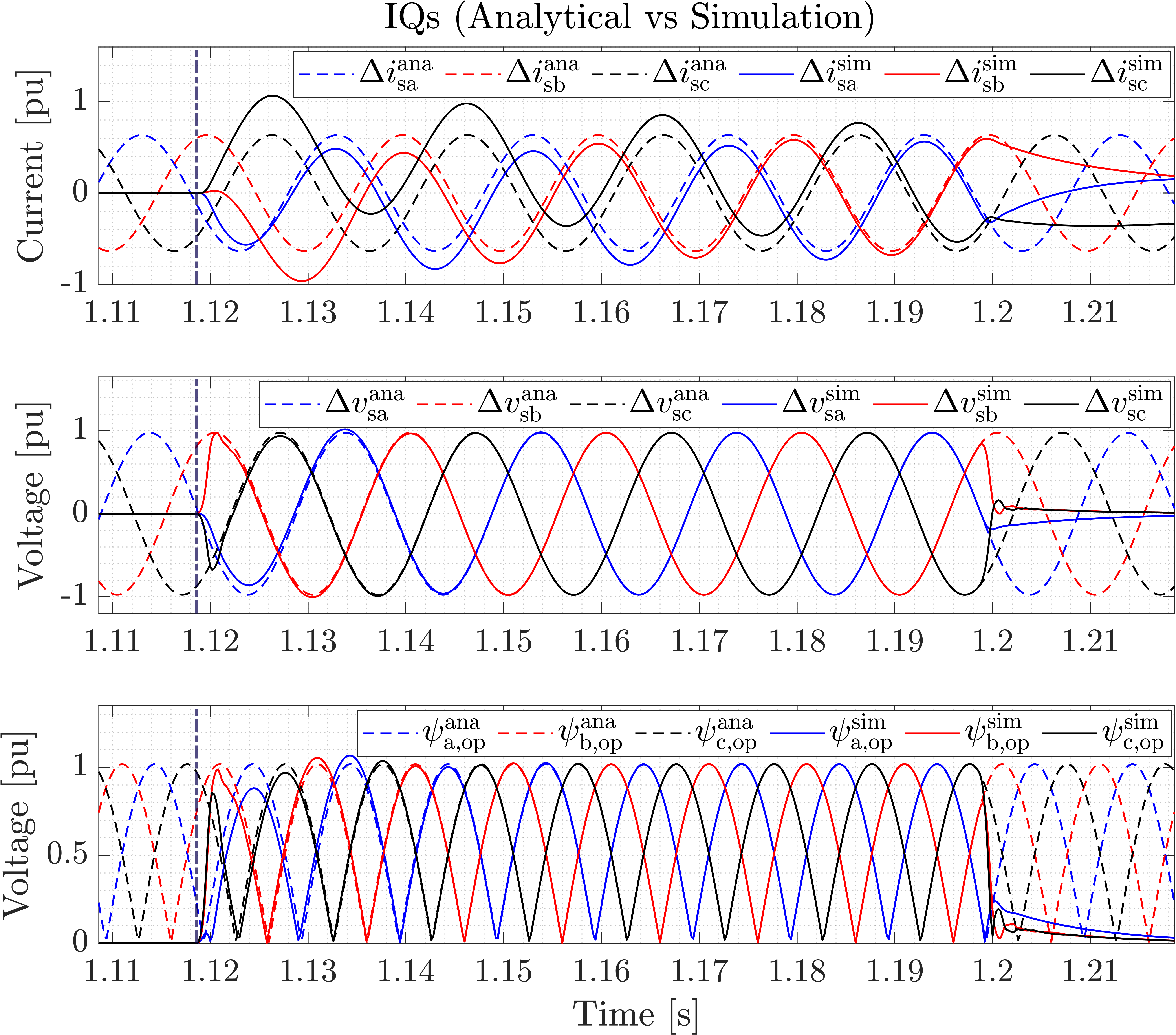}
    \caption{Analytical calculations (dashed lines) compared to PSCAD simulations (solid lines): $\Delta i_\mathrm{s}(t)$, $\Delta v_\mathrm{s}(t)$ and $\psi_\mathrm{}^\mathrm{op}(t)$ ($m_\mathrm{f}=0.2$ pu).}
    \label{fig:ABCG_20_10_0_60_50_5_psi_op}
\end{figure}

The test system in Fig.~\ref{fig:IQ-PSCAD-setup} was implemented in PSCAD to validate the proposed modelling of nonlinear source characteristics on the relay-measured IQs under balanced faults. It is the same system described in Section~\ref{sec:system-description}, but with a strong conventional source at the sending end. However, at $t=0$, the magnitude of the sending-end source impedance is increased by a factor of $5$, the magnitude of the internal voltage is decreased from $1$ pu to $0.33$ pu, and a balanced fault of $10$ $\Omega$ is applied at $m_\mathrm{f}=0.2$ pu. Fig.~\ref{fig:ABCG_20_10_0_60_50_5_psi_op} shows $\Delta i_\mathrm{s}(t)$, $\Delta v_\mathrm{s}(t)$ and $\psi^\mathrm{op}(t)$ obtained from the simulation, alongside with their steady-state analytically computed values from (\ref{IQ-I-phasor}), (\ref{IQ-V-phasor}) and (\ref{Distance-IQ-op-GFM}). The incremental voltages and currents have been computed using $p=2$ in (\ref{IQ-definition}) and have been passed through a third-order Butterworth low-pass filter with a cutoff frequency of $450$ Hz. As can be seen, the steady-state simulation results match the analytical calculations based on the proposed model in Fig.~\ref{fig:IQ-network-TD-and-phasor}. Moreover, $\Delta v_\mathrm{s}(t)$ quickly settles within only a few milliseconds to its steady-state waveform, unlike $\Delta i_\mathrm{s}(t)$, which has a decaying DC offset. The quick settling also applies to $\psi^\mathrm{op}(t)$ (it is also a voltage quantity), although around half a fundamental period ($10$ ms).

The results in Fig.~\ref{fig:ABCG_20_10_0_60_50_5_psi_op} correspond to only one of many simulations conducted to validated the proposed modelling in Fig.~\ref{fig:IQ-network-TD-and-phasor}. A Python script was coded to automate hundreds of PSCAD simulation runs on the system in Fig.~\ref{fig:IQ-PSCAD-setup}, where the magnitude and phase angle of the sending-end voltage source, the resistance and inductance of the sending-end source impedance, as well as the fault resistance and location, were varied. The derived expressions in (\ref{IQ-I-phasor}), (\ref{IQ-V-phasor}) and (\ref{Distance-IQ-op-GFM}) always matched perfectly with the obtained steady-state waveforms from the simulations. Moreover, $\Delta v_\mathrm{s}(t)$ and $\psi^\mathrm{op}(t)$ quickly adjust to new steady-state waveforms when stepping between different values of $R_\mathrm{f}(t)$, $\Delta R_\mathrm{s}(t)$, $\Delta L_\mathrm{s}(t)$ and $\Delta e_\mathrm{s}(t)$ in the same simulation run, if the memory function in (\ref{IQ-definition}) remains within the pre-disturbance network (initial load condition). Consequently, equation (\ref{Operating}) derived from Fig.~\ref{fig:IQ-network-TD-and-phasor}~(d), gives a quasi-static approximation of $\psi^\mathrm{op}(t)$, where each moment in time is approximated as a new steady-state equilibrium. From this perspective, (\ref{Operating}) can be used to evaluate the expected performance of IQ-based distance protection when the sending-end source is a weak-infeed GFM inverter. Additionally, $\psi^\mathrm{op}(t)$ develops from $0$ when a fault occurs, whereas $\psi^\mathrm{rst}(t)$ does not, making it difficult to determine if a fault is internal or external during the initial energization of $\psi^\mathrm{op}(t)$.

\section{Interoperability Evaluation of IQ-Based Distance Protection and Quadrilateral Distance Protection With GFM Inverters}
\label{sec:Interoperability-IQ-dist-GFM}
It becomes clear from the explanations in Section.~\ref{GFM-IQ-impact-explain}, that certain current limiters are easier and more intuitive to model with a variable source impedance and others with a change of the IVS, i.e. the direct models are more practical. Therefore, considering the current-limiting control strategies of GFM1 and GFM2, which will be used to validate the analytical analysis of IQ-based distance protection and quadrilateral distance protection algorithms, it is practical to model their nonlinear source characteristics as additional time-varying source impedances, using $\Delta R_\mathrm{s}(t)$ and $\Delta L_\mathrm{s}(t)$ ($\Delta e_\mathrm{s}(t)=0$). 

Parameter sweeps are conducted in MATLAB on the system shown in Fig.~\ref{fig:SLIB-system}, from which the sending-end incremental current (\ref{IQ-I-phasor}), incremental voltage (\ref{IQ-V-phasor}) and operating quantity used for IQ-based distance protection (\ref{Distance-IQ-op-GFM}) are calculated, as well as the non-IQ sending-end current, voltage and apparent loop impedance. The restraining quantity (\ref{Restraining}) is calculated with $K=1$. The initial system condition and parameters are the same as explained in Section~\ref{sec:system-description}, unless anything else is stated. Hence, the SIR of the GFM inverter output filter ($\mathrm{SIR}_\mathrm{s}$) and remote-end grid source ($\mathrm{SIR}_\mathrm{g}$) are $0.3$, and the pre-fault load is $P_\mathrm{pre}=1$ pu. 

Zone 1 and zone 2 of the quadrilateral distance element are tuned to work reliably with an SG under a wide variety of fault locations and resistances (more details are provided in \cite{HJ-GFM-distance}). The reach of zone 1 is $m=80\%$ of the total line length, and it is deemed to pick up a fault if the calculated steady-state apparent impedances settle within zone 1. For the IQ-based distance element, the running sum of a fault loop is activated the moment in time when it is greater than zero and is not allowed to go below zero. The reach of the IQ-based distance element is also set to $m=80\%$, and it has the possibility to trip if the steady-state $\psi^\mathrm{op}$ is greater than $\psi^\mathrm{rst}$, as explained in Section~\ref{sec:IQ-dist-trip-variables}. For a pre-defined trip threshold level on the running sums, a large difference yields a high dependability margin.

\subsection{Internal Fault Dependability}
\begin{figure}[h]
    \centering
    \includegraphics[width=1\linewidth]{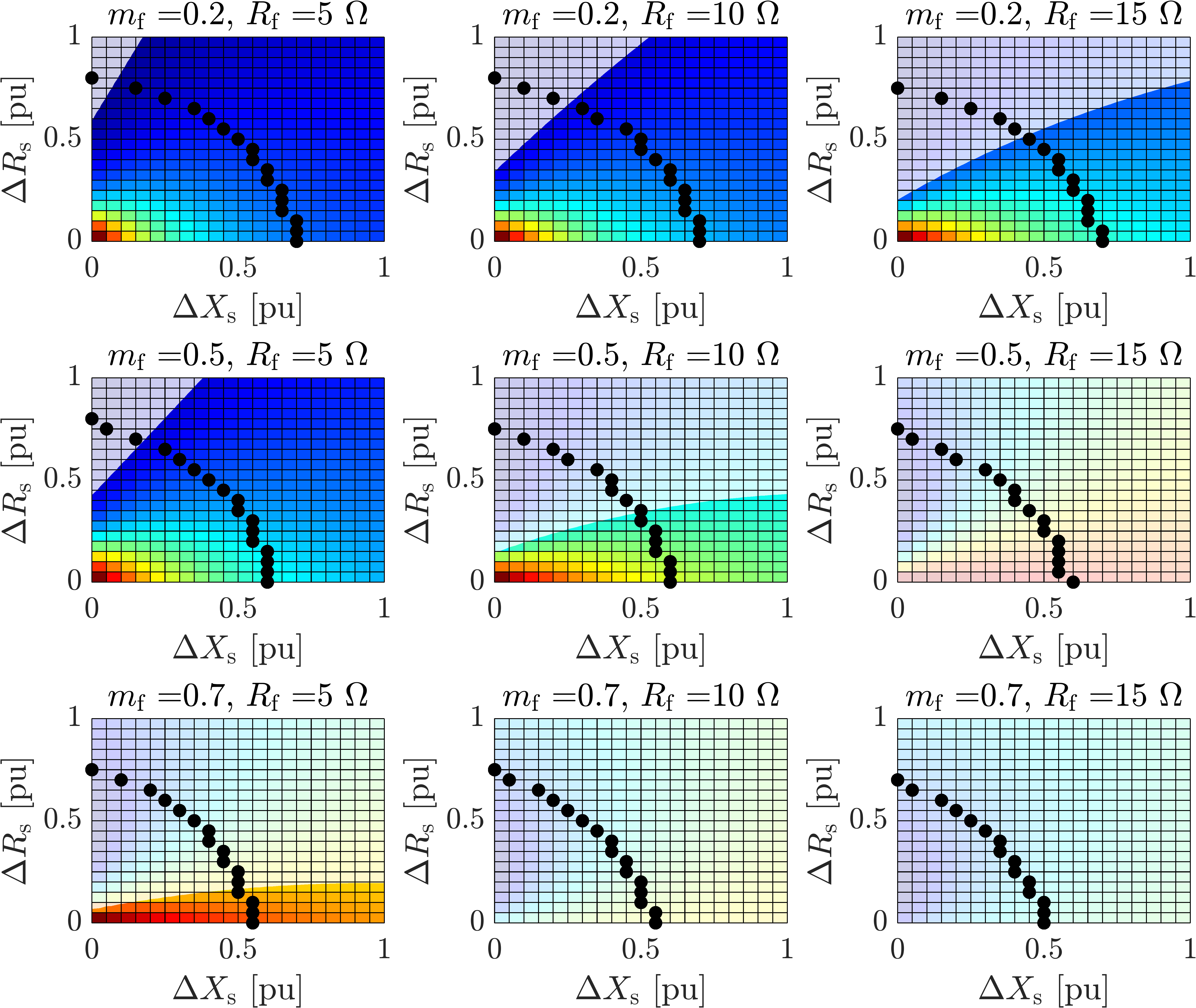}
    \caption{\textbf{IQ-based distance}: $\mathrm{SIR}_\mathrm{s}=\mathrm{SIR}_\mathrm{g}=0.3$ and $P_\mathrm{pre} = 1$ pu.}
    \label{fig:SIRg-0.3-SIRs-0.3-P3ph-1pu-LineLength-100km_absPsiOp}
\end{figure}

Fig.~\ref{fig:SIRg-0.3-SIRs-0.3-P3ph-1pu-LineLength-100km_absPsiOp} shows the performance of IQ-based distance protection with GFM inverters as function of $\Delta R_\mathrm{s}$ and $\Delta X_\mathrm{s}$, under various internal balanced faults. The shaded gray area in each subplot corresponds to the area where $\psi^\mathrm{op}\leq\psi^\mathrm{rst}$, meaning that IQ-based distance protection is not dependable. The colored areas indicate where $\psi^\mathrm{op}>\psi^\mathrm{rst}$, meaning the time-domain running sums (\ref{Running-sum-equation}) trend upwards and the distance element may pick up the fault. The black dots form the boundary where the magnitude of the injected current $|\bar{I}_\mathrm{s}|$ from the GFM inverter becomes $1.2$~pu (desired current limitation). 

It can be seen from Fig.~\ref{fig:SIRg-0.3-SIRs-0.3-P3ph-1pu-LineLength-100km_absPsiOp} that IQ-based distance protection is more dependable with GFM inverters if the current-limiting control strategy emulates a predominantly inductive source impedance change (for the same amount of overcurrent). The positive-sequence line angle is included in the small colored area of the subplot corresponding to $m_\mathrm{f}=0.7$ and $R_\mathrm{f}=5$~$\Omega$, implying that the running sums of the IQ-based distance element (\ref{Running-sum-equation}) trend upwards if the GFM inverter in Fig.~\ref{fig:SLIB-system} employs such a virtual-impedance-based current limiter to achieve $|\bar{I}_\mathrm{s}|=1.2$~pu, although slower than with an SG. On the other hand, if a saturation current limiter is used by the GFM inverter in Fig~\ref{fig:SLIB-system}, which emulates an adaptive virtual resistance ($\Delta L_\mathrm{s}=0$), the running sums (\ref{Running-sum-equation}) trend downwards when $|\bar{I}_\mathrm{s}|=1.2$~pu for all of the nine presented fault cases. Although current limitation is a dynamical process and the time-domain IQ-based distance element could pick up the fault before $\psi^\mathrm{op}\leq \psi^\mathrm{op}$ by having a sufficiently low trip threshold level on the running sums, such an implementation might jeopardize security against external faults while the IQ network is being energized or due to high superimposed oscillations.

\begin{figure}[h]
    \centering
    \includegraphics[width=1\linewidth]{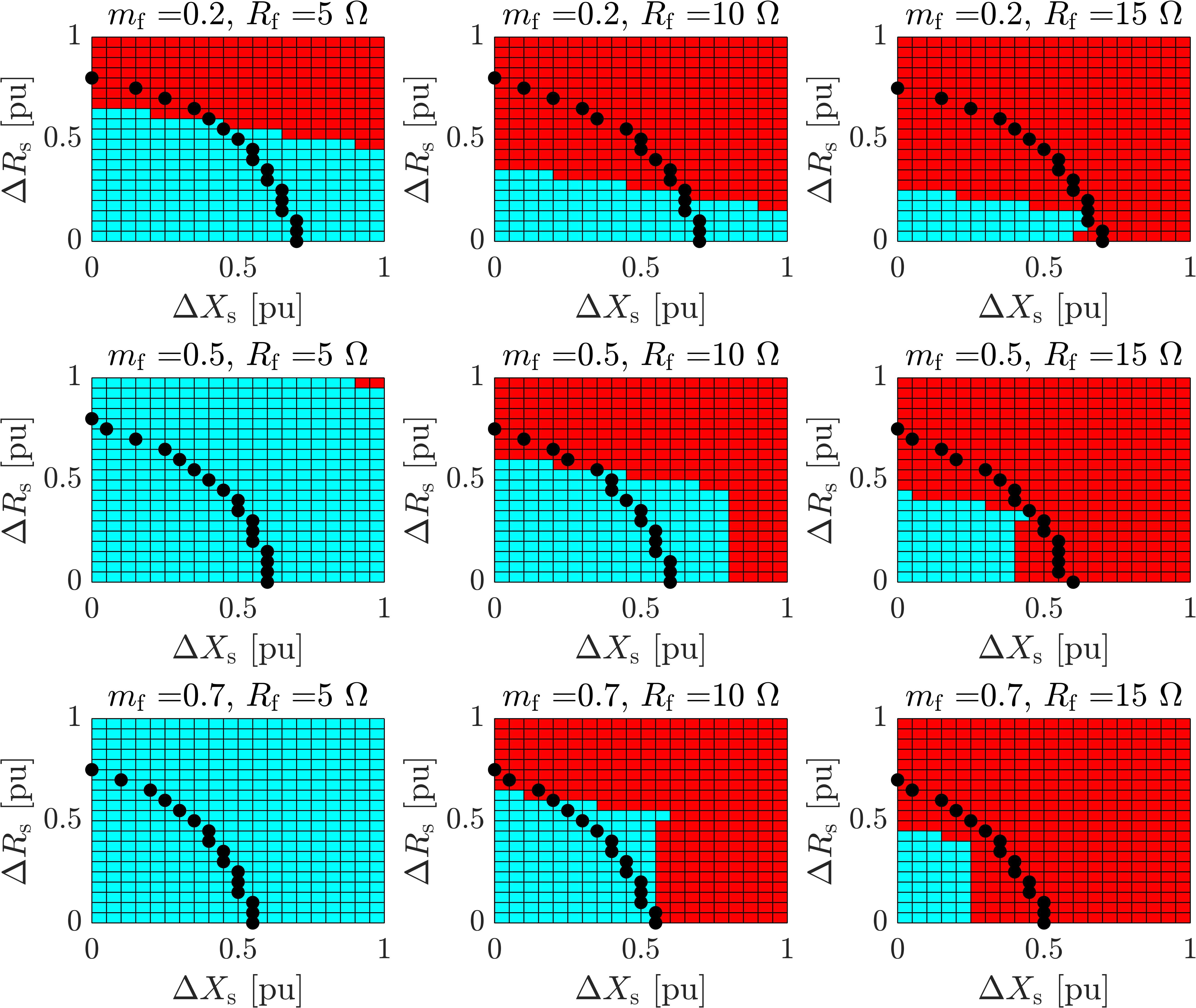}
    \caption{\textbf{Quadrilateral distance}: $\mathrm{SIR}_\mathrm{s}=\mathrm{SIR}_\mathrm{g}=0.3$ and $P_\mathrm{pre} = 1$ pu.}
    \label{fig:SIRg-0.3-SIRs-0.3-P3ph-1pu-LineLength-100km_Zsa_Logical}
\end{figure}

Fig.~\ref{fig:SIRg-0.3-SIRs-0.3-P3ph-1pu-LineLength-100km_Zsa_Logical} shows the performance of quadrilateral distance protection with GFM inverters as function of $\Delta R_\mathrm{s}$ and $\Delta X_\mathrm{s}$ under balanced faults. The cyan region in each subplot corresponds to when the final values of the apparent loop impedances settle within zone 1---for the red regions they settle outside. It can be seen that quadrilateral distance protection risks not picking up close-in balanced faults with GFM inverters, which is particularly true if the source impedance change to limit the injected current is not predominantly inductive. When comparing Fig.~\ref{fig:SIRg-0.3-SIRs-0.3-P3ph-1pu-LineLength-100km_absPsiOp} and \ref{fig:SIRg-0.3-SIRs-0.3-P3ph-1pu-LineLength-100km_Zsa_Logical}, it is seen that IQ-based distance protection demonstrates higher dependability for close-in faults, whereas quadrilateral distance protection is more dependable for faults occurring closer to the reach point. When $m_\mathrm{f}<0.2$, the quadrilateral distance relay struggles to be dependable during resistive faults, even when the current limiter of the GFM inverter emulates an inductive virtual impedance. Regarding solid balanced faults, both relays are dependable as function of $\Delta R_\mathrm{s}$ and $\Delta X_\mathrm{s}$, as well as the fault location $m_\mathrm{f}$.

\subsection{External Fault Security}
\begin{figure}[h]
    \centering
    \includegraphics[width=1\linewidth]{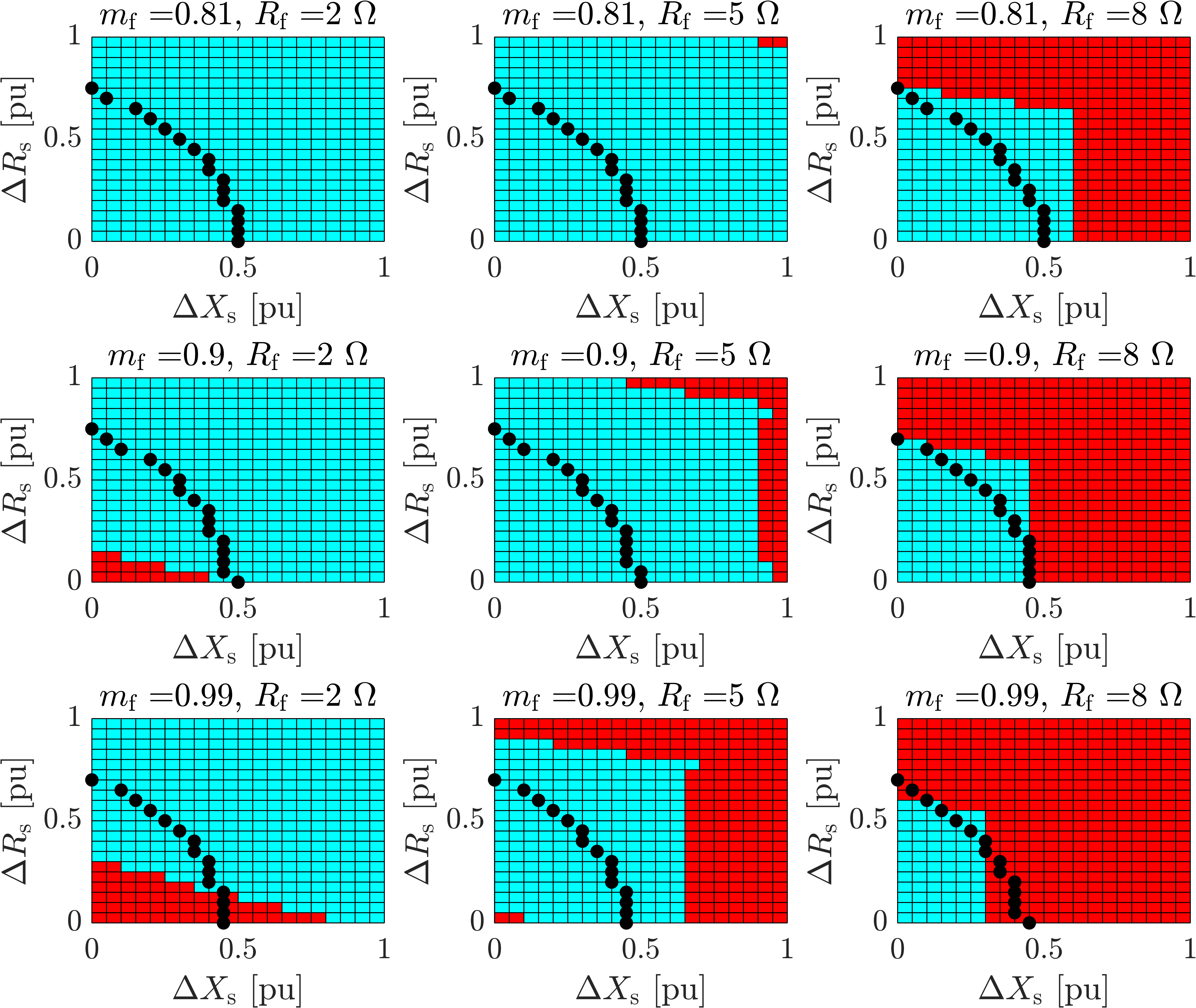}
    \caption{\textbf{Quadrilateral distance}: $\mathrm{SIR}_\mathrm{s}=\mathrm{SIR}_\mathrm{g}=0.3$ and $P_\mathrm{pre} = 1$ pu.}
    \label{fig:Zone2-LowRf-SIRg-0.3-SIRs-0.3-P3ph-1pu-LineLength-100km_Zsa_Logical}
\end{figure}

With regards to security against external faults, the analytical calculations always yield $\psi^\mathrm{op} \leq \psi^\mathrm{rst}$ as function of $\Delta R_\mathrm{s}$ and $\Delta L_\mathrm{s}$, even for balanced faults at $m_\mathrm{f}=0.81$ (from solid to high-resistance faults), indicating that IQ-based distance protection demonstrates credible security properties against external faults under GFM inverter current limitation. On the other hand, zone 1 of the quadrilateral distance element risks overreaching for external resistive faults, although it is secure under solid faults in the steady state. Fig.~\ref{fig:Zone2-LowRf-SIRg-0.3-SIRs-0.3-P3ph-1pu-LineLength-100km_Zsa_Logical} shows the security of the quadrilateral distance element as function of $\Delta R_\mathrm{s}$ and $\Delta L_\mathrm{s}$, where the cyan color means the computed apparent impedances settle inside zone 1, i.e. the relay trips and overreaches, whereas the red color means zone 1 is secure. As can be seen, the current limitation of GFM inverters can jeopardize the otherwise secure operation of quadrilateral distance protection from at least $m_\mathrm{f}=0.9$ and beyond, especially if it does not emulate a predominantly inductive virtual impedance. Of course, both IQ-based distance protection and quadrilateral distance protection zone 1 risk overreaching during the transient period of external faults; this depends on the design of the trip criterion and how fast the relays are allowed to trip. However, the steady-state overreach scenarios of quadrilateral distance protection from the analytical findings in Fig.~\ref{fig:Zone2-LowRf-SIRg-0.3-SIRs-0.3-P3ph-1pu-LineLength-100km_Zsa_Logical}, make such relays highly likely to overreach for those faults with GFM inverters, whereas time-domain IQ-based distance protection has the possibility to be designed to be secure.

\subsection{Proposed Trip Criterion for IQ-Based Distance Protection}

For time-domain IQ-based distance protection, dependability and security are heavily affected by where the trip threshold level of the running sums is put, which is not an easy-to-tune or straightforward user-specific setting that can be generalized for applications near GFM inverters, if it is desired to be both dependable and secure over a wide range of faults. This is because the running sums accumulate upwards at different slopes and superimposed oscillations, depending on the fault location and resistance, as well as the dynamical properties of the sending-end source, which is seen in Fig.~\ref{fig:SIRg-0.3-SIRs-0.3-P3ph-1pu-LineLength-100km_absPsiOp} (the origin corresponds to an SG). Hence, other means for tripping in time-domain IQ-based distance protection schemes are desired to facilitate easy settings that can be generalized to ensure dependable and secure distance protection with various nonlinear weak-infeed sources, such as GFM inverters with a limited overcurrent capability. 

Since the running sums produced by the GFM inverters have superimposed $100$~Hz oscillations, it is proposed to use a primary trip criterion based on the consecutive time that a running sum has been above $0$, such as $10$~ms or slightly longer, instead of a trip threshold level. The superimposed oscillations are produced due to the phase shift between $\psi^\mathrm{op}(t)$ and $\psi^\mathrm{rst}(t)$, but if $\psi^\mathrm{op}(t)$ has a higher magnitude than $\psi^\mathrm{rst}(t)$, the running sum of their difference will remain above $0$ for a longer consecutive time than the $10$~ms period of the superimposed oscillations (the opposite occurs if $\psi^\mathrm{op}<\psi^\mathrm{rst})$. This novel trip criterion is a fast time-domain approach to determine whether the magnitude of $\psi^\mathrm{op}(t)$ is higher than $\psi^\mathrm{rst}(t)$, shortly after the IQ network in Fig.~\ref{fig:IQ-network-TD-and-phasor}~(c) has been energized by the fault point, while also being independent of how steeply the running sums are accumulating upwards (the trend of the signals). Therefore, it is easier to generalize to multiple types of sources under various fault locations and resistances, while also being able to process nonlinear data. In fact, with both GFM1 and GFM2, the running sums can reach similar values for various internal and external faults due to the superimposed $100$~Hz oscillations, although slowly trending upwards for internal faults. A threshold level cannot be used to discriminate between internal and external faults in such cases, but the proposed trip criterion can. 

\begin{figure}[h]
    \centering
    \includegraphics[width=0.95\linewidth]{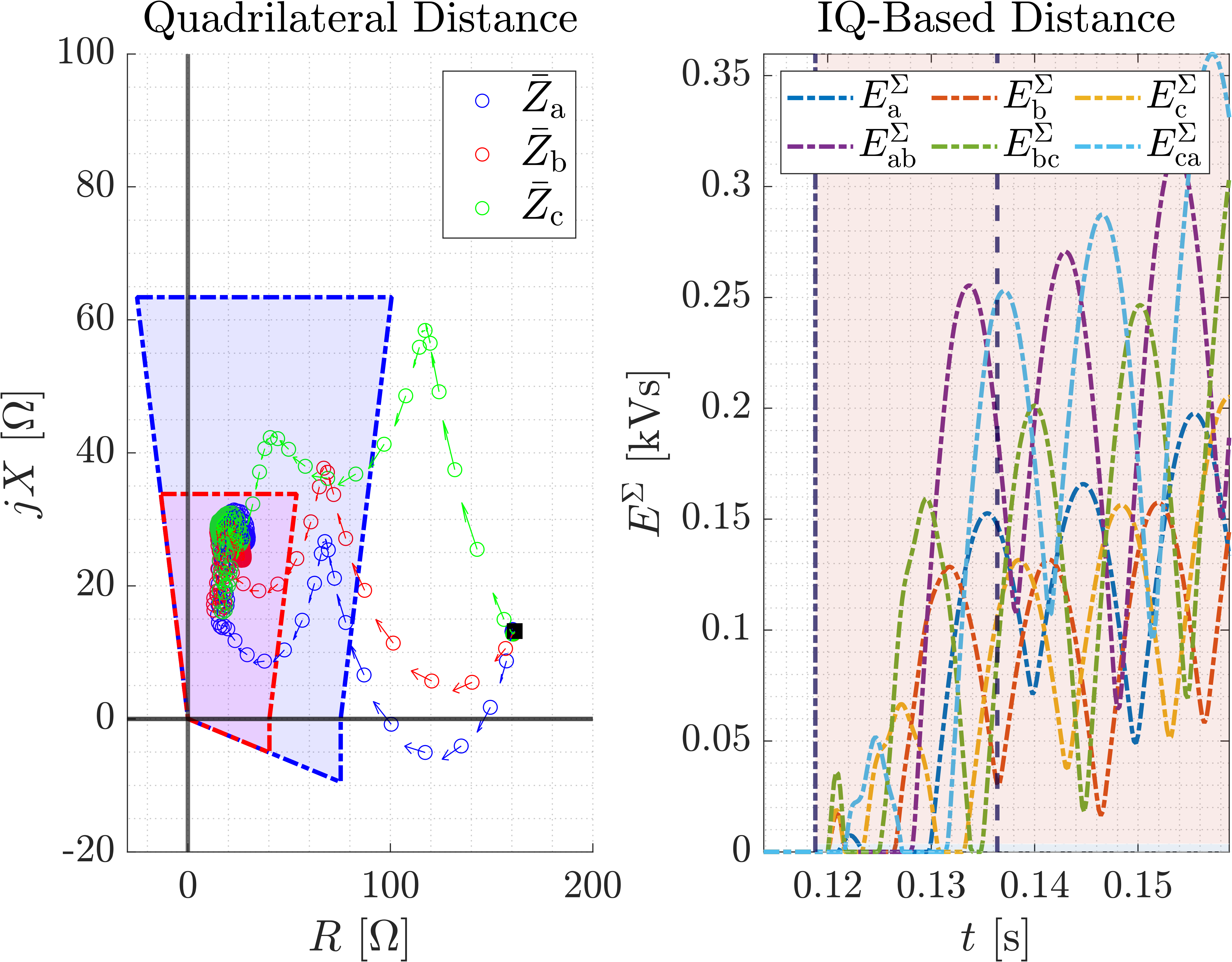}
    \caption{\textbf{GFM2}: ABCG fault at $m_\mathrm{f} = 70\%$ with $R_\mathrm{f} = 5$ $\Omega$.}
    \label{fig:GFM2-ABCG-70-5-0-1-dist-plot}
\end{figure}

To verify the analytical findings and the proposed trip criterion, PSCAD simulations of GFM1, GFM2 and an SG are conducted, as explained in Section.~\ref{sec:system-description}. The operations of the quadrilateral and IQ-based distance elements from the simulations match well with the analytical findings. Their sampling frequency is $5$~kHz and $p=2$ in (\ref{IQ-definition}). Fig.~\ref{fig:GFM2-ABCG-70-5-0-1-dist-plot} shows the performance of the distance elements with GFM2 when $m_\mathrm{f}=0.7$ and $R_\mathrm{f}=5$~$\Omega$. The apparent impedance scattered data are plotted with $1$~ms time increments and the running sums are not allowed to go below zero. As can be seen, the apparent impedances settle inside zone 1 and the running sums slowly trend upwards; this is in agreement with the results from Fig.~\ref{fig:SIRg-0.3-SIRs-0.3-P3ph-1pu-LineLength-100km_absPsiOp} and \ref{fig:SIRg-0.3-SIRs-0.3-P3ph-1pu-LineLength-100km_Zsa_Logical}. The rightmost dashed line shows when the trip signal is issued by the IQ-based distance element. With an SG or less current limitation by GFM2, the running sums trend upwards faster, but the trip speed remains almost unchanged.

\begin{figure}[h]
    \centering
    \includegraphics[width=0.95\linewidth]{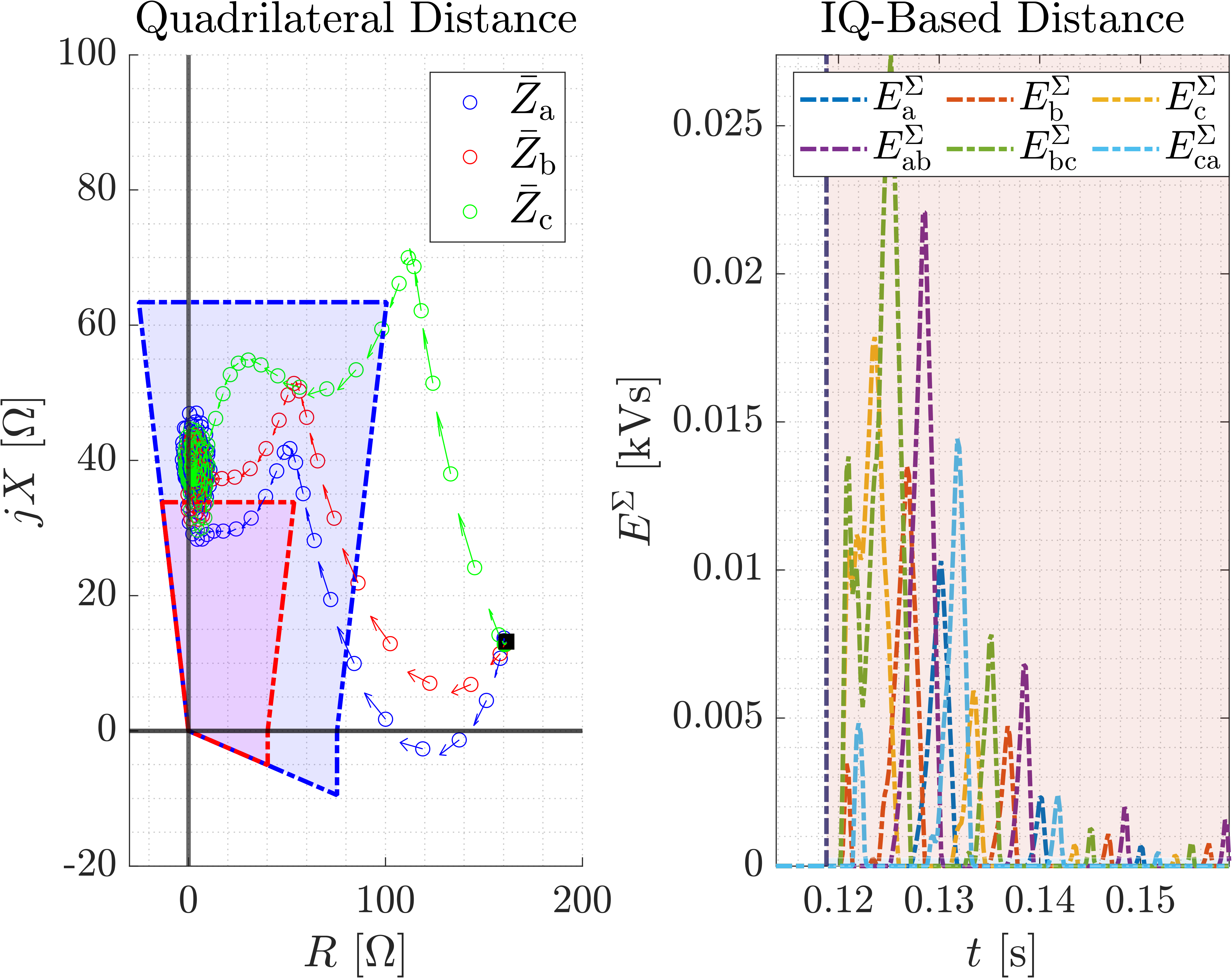}
    \caption{\textbf{GFM2}: ABCG fault at $m_\mathrm{f} = 90\%$ with $R_\mathrm{f} = 0$ $\Omega$.}
    \label{fig:GFM2-ABCG-90-0-0-1-dist-plot}
\end{figure}

Out of the simulated faults, the proposed IQ-based distance element is dependable with GFM2 up to $R_\mathrm{f}=15$~$\Omega$ at $m_\mathrm{f}=0.45$, $R_\mathrm{f}=10$~$\Omega$ at $m_\mathrm{f}=0.6$, and $R_\mathrm{f}=8$~$\Omega$ at $m_\mathrm{f}=0.75$. The slightly enhanced dependability from the simulations compared to the analytical findings in Fig.~\ref{fig:SIRg-0.3-SIRs-0.3-P3ph-1pu-LineLength-100km_absPsiOp}, is due to the current limitation not being instantaneous. It is also secure for all the simulated faults at $m_\mathrm{f}=0.9$ and beyond, whereas the quadrilateral distance element overreaches for most of those faults. Fig.~\ref{fig:GFM2-ABCG-90-0-0-1-dist-plot} demonstrates the security of the proposed IQ-based distance element when $R_\mathrm{f}=0$ and $m_\mathrm{f}=0.9$, whereas the quadrilateral distance element risks overreaching during the transient period (for higher $R_\mathrm{f}$, it overreaches in steady state). The dependability of the quadrilateral distance element with GFM2 matches the results in Fig.~\ref{fig:SIRg-0.3-SIRs-0.3-P3ph-1pu-LineLength-100km_Zsa_Logical}. Overall, the performance (trip or no trip) of the proposed IQ-based distance element with GFM2 is near identical to that with an SG---a feature that the quadrilateral distance element does not share.

\begin{figure}[h]
    \centering
    \includegraphics[width=0.95\linewidth]{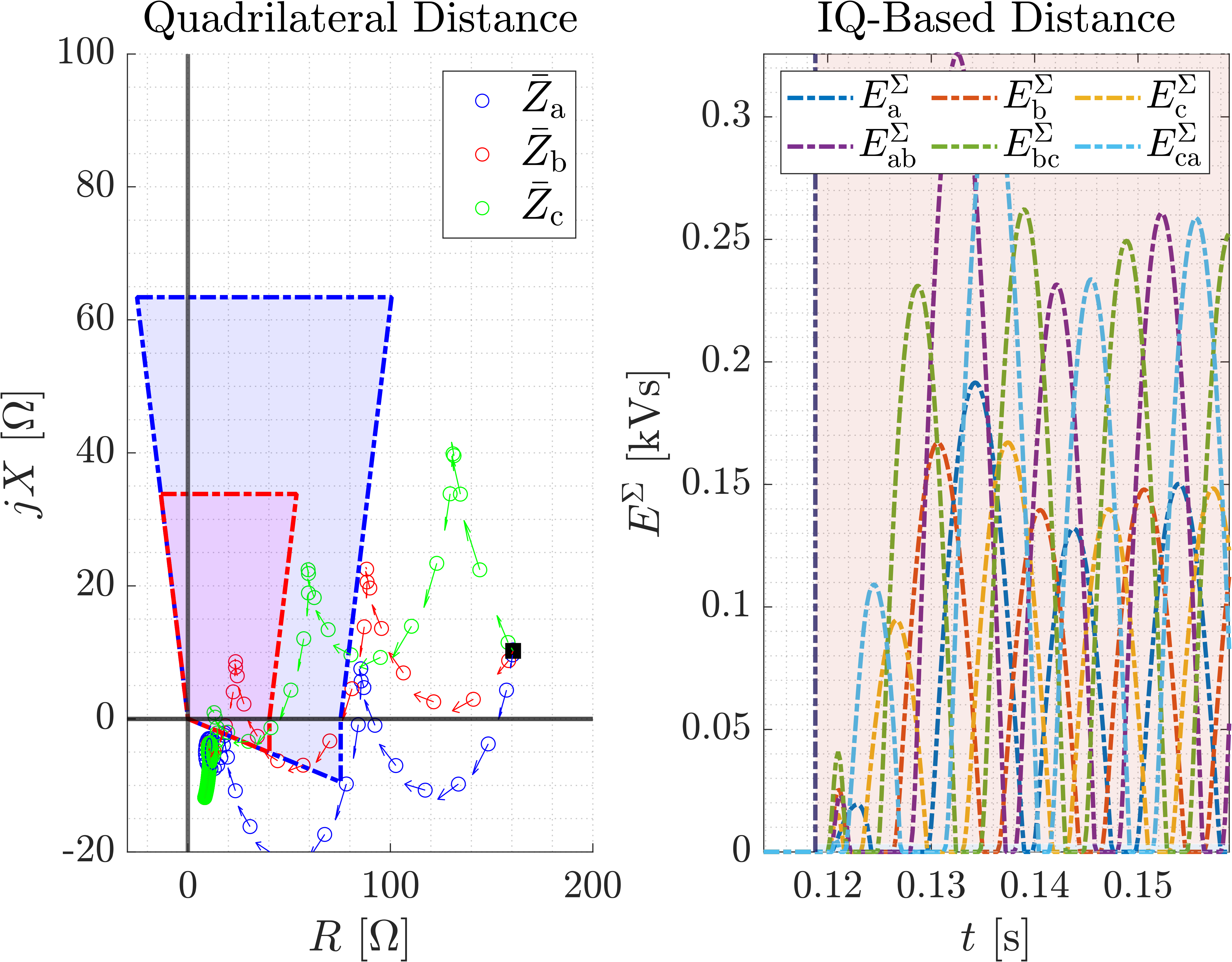}
    \caption{\textbf{GFM1}: ABCG fault at $m_\mathrm{f} = 50\%$ with $R_\mathrm{f} = 10$ $\Omega$.}
    \label{fig:GFM1-ABCG-50-10-0-1-dist-plot}
\end{figure}

With regards to GFM1, the proposed IQ-based distance element is dependable up to $R_\mathrm{f}=8$~$\Omega$ at $m_\mathrm{f}=0.2$, and $R_\mathrm{f}=5$~$\Omega$ at $m_\mathrm{f}=0.7$ (better than quadrilateral for close-in faults but worse near the reach point). Similar to GFM2, the slightly improved dependability compared to the analytical results in Fig.~\ref{fig:SIRg-0.3-SIRs-0.3-P3ph-1pu-LineLength-100km_absPsiOp}, is due to the current-limiting control dynamics not being instantaneous. Fig.~\ref{fig:GFM1-ABCG-50-10-0-1-dist-plot} shows the misoperation of both distance elements with GFM1, when $m_\mathrm{f}=0.5$ and $R_\mathrm{f}=10$~$\Omega$, which is in agreement with the analytical findings in Fig.~\ref{fig:SIRg-0.3-SIRs-0.3-P3ph-1pu-LineLength-100km_absPsiOp} and \ref{fig:SIRg-0.3-SIRs-0.3-P3ph-1pu-LineLength-100km_Zsa_Logical}. As for security, the proposed IQ-based distance element is secure for the simulated faults at $m_\mathrm{f}=0.9$ and beyond, except $R_\mathrm{f}=2$~$\Omega$, which is still a significant improvement compared to that of the quadrilateral element.

\section{Conclusion}
\label{sec:Conclusion}

This paper proposes a general analytical framework to model the impact of GFM inverters on the relay-measured IQs, uses it to investigate the interoperability of time-domain IQ-based distance protection in the vicinity of GFM inverters with different current-limiting control strategies, and compares the relay performance with that of quadrilateral distance protection. The analytical findings are validated with PSCAD simulations. Results show that time-domain IQ-based distance protection has the possibility to demonstrate superior dependability for close-in faults compared to quadrilateral distance protection with GFM inverters, as well as the possibility to be secure for external faults when quadrilateral distance protection overreaches. However, this desired performance is hard to achieve with various GFM inverters, due to the settings for tripping being sensitive to the GFM inverter FRT control. Taking the observed interoperability issues into account, a novel trip criterion for dependable and secure time-domain IQ-based distance protection is proposed, where time variables are used for tripping based on the characteristics of the running sum signals, instead of a trip threshold level. It facilitates easy-to-tune and straightforward settings that can be generalized for applications with various sources, including GFM inverters with different current-limiting control strategies. The novel trip criterion is motivated from the analytical findings and validated with PSCAD simulations of two vastly different GFM inverter models.

\section*{Acknowledgments}
The authors of this paper would like to thank Hitachi Energy and The Swedish Energy Agency who have financially supported this work.

%\bibliographystyle{IEEEtran}
%\bibliography{bibliography.bib}

\begin{thebibliography}{10}
\providecommand{\url}[1]{#1}
\csname url@samestyle\endcsname
\providecommand{\newblock}{\relax}
\providecommand{\bibinfo}[2]{#2}
\providecommand{\BIBentrySTDinterwordspacing}{\spaceskip=0pt\relax}
\providecommand{\BIBentryALTinterwordstretchfactor}{4}
\providecommand{\BIBentryALTinterwordspacing}{\spaceskip=\fontdimen2\font plus
\BIBentryALTinterwordstretchfactor\fontdimen3\font minus \fontdimen4\font\relax}
\providecommand{\BIBforeignlanguage}[2]{{%
\expandafter\ifx\csname l@#1\endcsname\relax
\typeout{** WARNING: IEEEtran.bst: No hyphenation pattern has been}%
\typeout{** loaded for the language `#1'. Using the pattern for}%
\typeout{** the default language instead.}%
\else
\language=\csname l@#1\endcsname
\fi
#2}}
\providecommand{\BIBdecl}{\relax}
\BIBdecl

\bibitem{Protection-100}
U.~Münz, S.~Bhela, N.~Xue, A.~Banerjee, M.~Reno, D.~Kelly, E.~Farantatos, A.~Haddadi, D.~Ramasubramanian, and A.~Banaie, ``Protection of 100\% inverter-dominated power systems with grid-forming inverters and protection relays - gap analysis and expert interviews,'' 04 2024.

\bibitem{IBR-impact-Haddadi}
\BIBentryALTinterwordspacing
A.~Haddadi, E.~Farantatos, I.~Kocar, and U.~Karaagac, ``Impact of inverter based resources on system protection,'' \emph{Energies}, vol.~14, no.~4, 2021. [Online]. Available: \url{https://www.mdpi.com/1996-1073/14/4/1050}
\BIBentrySTDinterwordspacing

\bibitem{Distance-unconventional}
B.~Kasztenny, ``Line distance protection near unconventional energy sources,'' in \emph{16th International Conference on Developments in Power System Protection (DPSP 2022)}, vol. 2022, 2022, pp. 224--229.

\bibitem{IBR-impact-Juan}
J.~C. Quispe and E.~Orduña, ``Transmission line protection challenges influenced by inverter-based resources: A review,'' \emph{Protection and Control of Modern Power Systems}, vol.~7, no.~3, pp. 1--17, 2022.

\bibitem{IBR-problems-Normann}
R.~Chowdhury and N.~Fischer, ``Transmission line protection for systems with inverter-based resources – part i: Problems,'' \emph{IEEE Transactions on Power Delivery}, vol.~36, no.~4, pp. 2416--2425, 2021.

\bibitem{GFM-ENTSO-E}
``High penetration of power electronic interfaced power sources and the potential contribution of grid forming converters,'' European Network of Transmission System Operators for Electricity, Tech. Rep., 2020.

\bibitem{IBR-impact-2025}
Z.~Yang, H.~Wang, W.~Liao, C.~L. Bak, and Z.~Chen, ``Protection challenges and solutions for ac systems with renewable energy sources: A review,'' \emph{Protection and Control of Modern Power Systems}, vol.~10, no.~1, pp. 18--39, 2025.

\bibitem{HJ-GFM-distance}
\BIBentryALTinterwordspacing
H.~Johansson, Q.~Xing, N.~Taylor, and X.~Wang, ``Impacts of grid-forming inverters on distance protection,'' \emph{IET Generation, Transmission \& Distribution}, vol.~19, no.~1, p. e13354, 2025. [Online]. Available: \url{https://ietresearch.onlinelibrary.wiley.com/doi/abs/10.1049/gtd2.13354}
\BIBentrySTDinterwordspacing

\bibitem{Bergeron-protection-GFM}
C.~L. Peralta and H.~N.~V. Pico, ``A time-domain protection approach for ac transmission systems with grid-forming resources,'' \emph{IEEE Transactions on Instrumentation and Measurement}, vol.~73, pp. 1--18, 2024.

\bibitem{Collaborative-distance-IBR}
C.~Chao, X.~Zheng, Y.~Weng, Z.~Liu, H.~Ye, H.~Liu, H.~Zhang, Y.~Liu, Y.~Wang, and N.~Tai, ``Collaborative solution of distance protection and dual current control for outgoing lines of inverter-based resources during line-to-line faults,'' \emph{IEEE Transactions on Smart Grid}, vol.~15, no.~4, pp. 3782--3794, 2024.

\bibitem{IQ-Naidu-souce-impedance}
N.~George, O.~D. Naidu, and A.~K. Pradhan, ``Distance protection for lines connecting converter interfaced renewable power plants: Adaptive to grid-end structural changes,'' \emph{IEEE Transactions on Power Delivery}, vol.~38, no.~3, pp. 2011--2021, 2023.

\bibitem{IQ-Naidu}
O.~D. Naidu, N.~George, S.~Zubic, and M.~Krakowski, ``Time-domain-based distance protection for transmission networks: Secure and reliable solution for complex networks,'' \emph{IEEE Access}, vol.~11, pp. 104\,656--104\,675, 2023.

\bibitem{Source-agnostic-double-line}
N.~George, O.~D. Naidu, and A.~Kumar~Pradhan, ``Source-agnostic single-ended protection and fault location for double-circuit lines connected to power electronics-based sources,'' \emph{IEEE Access}, vol.~13, pp. 132\,038--132\,051, 2025.

\bibitem{GFM-distance-Baeckeland}
N.~Baeckeland, D.~Venkatramanan, S.~Dhople, and M.~Kleemann, ``On the distance protection of power grids dominated by grid-forming inverters,'' in \emph{2022 IEEE PES Innovative Smart Grid Technologies Conference Europe (ISGT-Europe)}, 2022, pp. 1--6.

\bibitem{GFM-distance-Booth}
D.~Liu, Q.~Hong, M.~A.~U. Khan, A.~Dyśko, A.~E. Alvarez, and C.~Booth, ``Evaluation of grid-forming converter's impact on distance protection performance,'' in \emph{16th International Conference on Developments in Power System Protection (DPSP 2022)}, vol. 2022, 2022, pp. 285--290.

\bibitem{IQ-Vermunicht}
\BIBentryALTinterwordspacing
J.~Vermunicht, W.~Leterme, and D.~{Van Hertem}, ``Analysing the performance of incremental quantity based directional time-domain protection near {HVAC} cables and {VSC} {HVDC} converters,'' \emph{Electric Power Systems Research}, vol. 223, p. 109599, 2023. [Online]. Available: \url{https://www.sciencedirect.com/science/article/pii/S0378779623004881}
\BIBentrySTDinterwordspacing

\bibitem{LCD-IQ-transient-energy}
N.~George, O.~Naidu, and A.~K. Pradhan, ``Differential protection for lines connected to inverter-based resources: Problems and solution,'' in \emph{2022 22nd National Power Systems Conference (NPSC)}, 2022, pp. 419--424.

\bibitem{DEA-Type3}
O.~D. Naidu, N.~George, and V.~Pradhan, ``Distance protection for lines connected with type iii wind farms: Problems and solution,'' in \emph{2023 IEEE PES 15th Asia-Pacific Power and Energy Engineering Conference (APPEEC)}, 2023, pp. 1--6.

\bibitem{HJ-DEA-HongKong}
H.~Johansson, Q.~Xing, N.~Taylor, and X.~Wang, ``Differential equation algorithms for distance relays in the presence of inverter-based resources,'' in \emph{18th International Conference on Developments in Power System Protection (DPSP APAC 2025)}, vol. 2025, 2025, pp. 211--221.

\bibitem{IQ-real-world-faults}
E.~O. Schweitzer, B.~Kasztenny, and M.~V. Mynam, ``Performance of time-domain line protection elements on real-world faults,'' in \emph{2016 69th Annual Conference for Protective Relay Engineers (CPRE)}, 2016, pp. 1--17.

\bibitem{IQ-circle}
E.~O. Schweitzer and B.~Kasztenny, ``Distance protection: Why have we started with a circle, does it matter, and what else is out there?'' in \emph{2018 71st Annual Conference for Protective Relay Engineers (CPRE)}, 2018, pp. 1--19.

\bibitem{IQ-Tanbhir}
M.~T. Hoq, J.~Wang, and N.~Taylor, ``An incremental quantity based distance protection with capacitor voltage estimation for series compensated transmission lines,'' \emph{IEEE Access}, vol.~9, pp. 164\,493--164\,502, 2021.

\bibitem{IQ-park}
Y.~Chen, M.~Wen, Z.~Wang, X.~Yin, and D.~Chen, ``An incremental quantities distance protection using park's transformation,'' \emph{IEEE Transactions on Power Delivery}, vol.~37, no.~4, pp. 3200--3212, 2022.

\bibitem{HJ-IQ-Bilbao}
H.~Johansson, Q.~Xing, N.~Taylor, and X.~Wang, ``Incremental quantities protection schemes with grid-forming inverters,'' in \emph{19th IET Conference on Developments in Power System Protection (DPSP Europe 2025)}, vol. 2025, 2025, pp. 277--289.

\bibitem{IQ-IBR-impact}
H.~S. Samkari and B.~K. Johnson, ``Impact of distributed inverter-based resources on incremental quantities-based protection,'' in \emph{2021 IEEE Power \& Energy Society General Meeting (PESGM)}, 2021, pp. 1--5.

\bibitem{Jianping-Bilbao}
J.~Wang, Y.~Li, M.~Tajdinian, and M.~T. Hoq, ``New phase selection solution for distance protection with different system conditions,'' in \emph{19th IET Conference on Developments in Power System Protection (DPSP Europe 2025)}, vol. 2025, 2025, pp. 217--221.

\bibitem{SIR-tutorial}
M.~J. Thompson and A.~Somani, ``A tutorial on calculating source impedance ratios for determining line length,'' in \emph{2015 68th Annual Conference for Protective Relay Engineers}, 2015, pp. 833--841.

\bibitem{GFM-current-limitation-Xiongfei}
B.~Fan, T.~Liu, F.~Zhao, H.~Wu, and X.~Wang, ``A review of current-limiting control of grid-forming inverters under symmetrical disturbances,'' \emph{IEEE Open Journal of Power Electronics}, vol.~3, pp. 955--969, 2022.

\bibitem{GFM-current-limitation-Baeckeland}
N.~Baeckeland, D.~Chatterjee, M.~Lu, B.~Johnson, and G.-S. Seo, ``Overcurrent limiting in grid-forming inverters: A comprehensive review and discussion,'' \emph{IEEE Transactions on Power Electronics}, vol.~39, no.~11, pp. 14\,493--14\,517, 2024.

\bibitem{GFM-limiter-sat-resistive}
B.~Fan and X.~Wang, ``Equivalent circuit model of grid-forming converters with circular current limiter for transient stability analysis,'' \emph{IEEE Transactions on Power Systems}, vol.~37, no.~4, pp. 3141--3144, 2022.

\end{thebibliography}

% Generated by IEEEtran.bst, version: 1.14 (2015/08/26)

\vspace{11pt}

\vfill

\end{document}